\documentclass{huawei}

\usepackage{amssymb} 

\usepackage[T1]{fontenc}
\usepackage{newpxtext,newpxmath}
\usepackage{microtype}
\usepackage{setspace}
\usepackage{xcolor}

\usepackage[most]{tcolorbox}
\usepackage[table]{xcolor}

\usepackage{xeCJK}

\definecolor{deepblue}{RGB}{37,99,235}
\definecolor{lightblue}{RGB}{239,246,255}

\usepackage[utf8]{inputenc} 
\usepackage[T1]{fontenc}    
\usepackage{hyperref}       
\usepackage{url}            
\usepackage{booktabs}       
\usepackage{amsfonts}       
\usepackage{nicefrac}       
\usepackage{microtype}      
\usepackage{fancyhdr}  
\usepackage{graphicx}
\usepackage{soul}

\usepackage{multirow}
\usepackage{multicol}
\usepackage{makecell}
\usepackage{arydshln}

\usepackage{enumitem}
\usepackage{xcolor}

\usepackage{graphicx}
\usepackage{fancyhdr}
\usepackage{xcolor}

\begin{document}

\title{
SearchArt: Training Long-Horizon Search Agent with Scalable Synthetic and Verified Tasks
\vspace{-3mm}
}

%

\author{
Huawei Cloud Post-Training Team\footnote{We welcome discussion, collaboration, and contributions to SearchArt: chenchong55@huawei.com, meilang1@huawei.com}
\vspace{-5mm}
}





\abstract{
Recent advances in large language models (LLMs) have enabled search agents to autonomously tackle complex tasks across extended search and reasoning horizons.
However, training effective search agents remains challenging due to the lack of scalable and long-horizon tasks, 
and the difficulty of evaluating and correcting intermediate reasoning and tool-use behaviors.
We introduce \textbf{SearchArt}, a scalable framework for training long-horizon search agents through verification-driven task synthesis and a multi-stage post-training pipeline.
SearchArt constructs large-scale datasets for complex \textbf{search}-, \textbf{research}- and \textbf{user}-oriented tasks by synthesizing diverse information-seeking QA pairs and corresponding search trajectories from web documents and automatically generated evidence graphs.
To ensure the reliability of the synthesized data, we design a verification pipeline that jointly evaluates QA consistency, trajectory quality, and the relevance of retrieved evidence. The verified trajectories are subsequently used in a multi-stage training process comprising supervised fine-tuning and reinforcement learning-based policy optimization.
Search agents trained with SearchArt exhibit adaptive search planning, iterative evidence aggregation, and complex reasoning over extended interaction horizons. 
Experimental results demonstrate that, with only (Qwen3.5-) 27B parameters, SearchArt scores \textbf{74.39 on BrowseComp-ZH}, \textbf{70.06 on BrowseComp}, and \textbf{52.55 on Deepresearch-bench}, matching or surpassing frontier closed-source agents on both deepsearch and deepresearch benchmarks.
}

\maketitle
\vspace{-8mm}

\begin{figure}[htbp]
    \centering
    \includegraphics[width=0.96\linewidth]{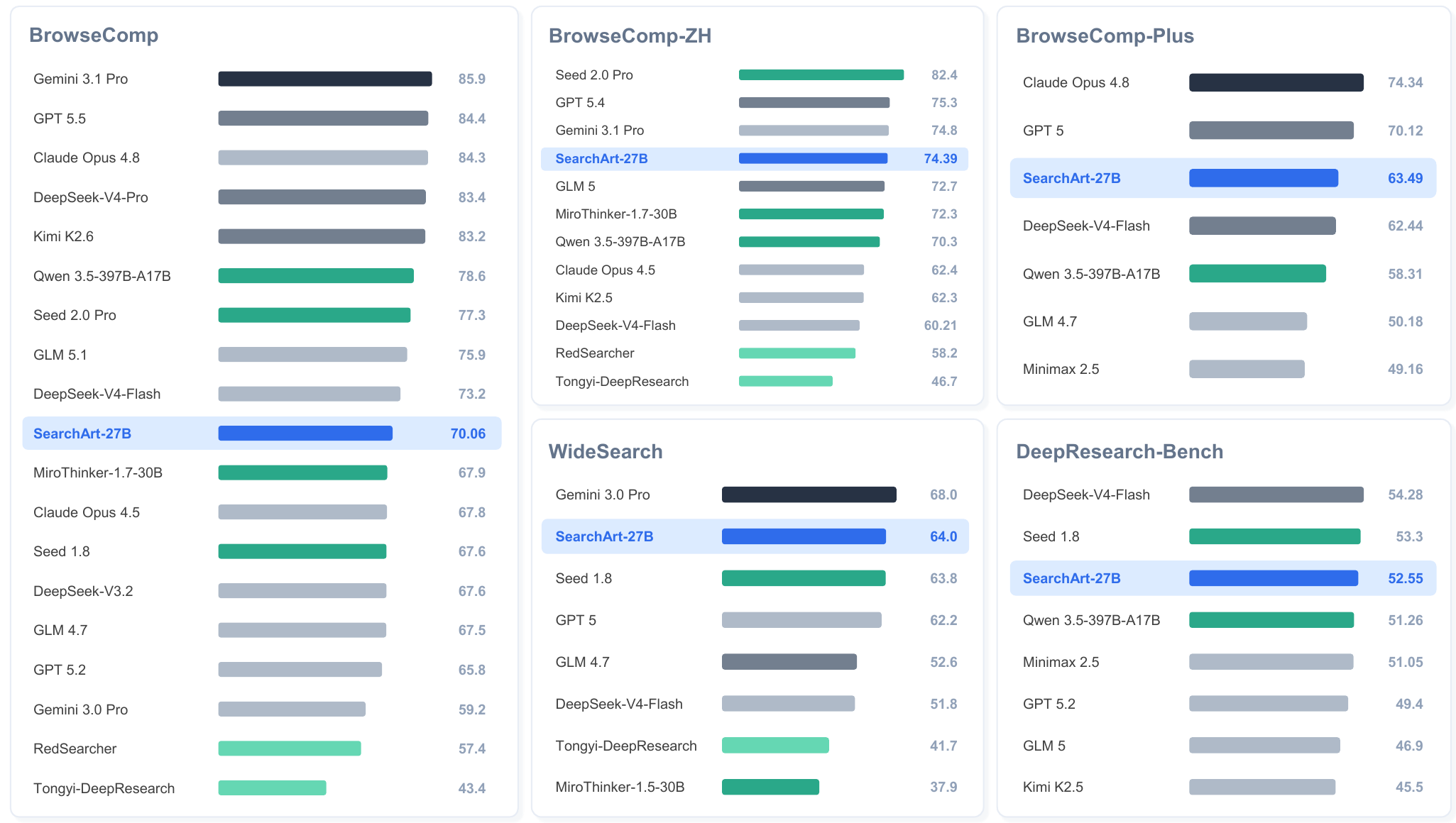}
    \caption{Comprehensive comparison across five benchmarks that evaluate different capabilities.}
    \label{fig:browsecomp_leaderboard_overview}
\end{figure}

\tableofcontents
\newpage

\section{Introduction}
Large language models (LLMs) \cite{naveed2025comprehensive, zhao2023survey, achiam2023gpt, qwenbai2023qwen, liu2024deepseek, guo2025deepseek, xu2026deepseek, yin2024survey} are increasingly capable of acting as autonomous agents for information retrieval and synthesis, enabling them to solve complex tasks through multi-turn interactions.
Unlike conventional question-answering (QA) systems \cite{lewis2020retrieval, gao2023retrieval, mei2025survey, liu2025llm4ranking, gong2026cardrewriter, niu2025distributionally, mei2025dense, niu2025addressing, mei2023improving}, that operate over a fixed context, long-horizon search agents \cite{team2026mirothinker, chu2026redsearcher, deng2026fort, du2026openseeker, zhu2026marco, hu2025step} must dynamically formulate and revise retrieval strategies, maintain evolving representations of the available evidence, and integrate information from heterogeneous sources throughout extended reasoning processes. These capabilities are critical for research-intensive tasks, including literature analysis, comparative studies, scientific discovery, and strategic decision-making.

Although agentic reasoning \cite{singh2025agentic, wu2025agentic, wu2025agentic1} has advanced rapidly in recent years, training effective search agents \cite{team2026mirothinker, chu2026redsearcher, jin2025search, song2025r1, yu2026m, mei2025ai, xia2026search, gong2024cosearchagent, chen2026unitymas, yang2026tournament, zhang2026oases, chen2025mao, chen2026beyond, chen2026jade, chen2026improving} remains a fundamental challenge. Despite the growing capabilities of LLMs, existing approaches continue to be limited by the scarcity of high-quality QA data, the substantial cost of human annotation, and the difficulty of evaluating and correcting intermediate reasoning and tool-use behaviors.
These limitations hinder scalability, particularly for long-horizon research tasks that require diverse retrieval strategies, complex multi-step reasoning, and reliable attribution of supporting evidence.
Moreover, collected reasoning trajectories often vary considerably in quality, while supervisory signals for long-horizon reasoning remain sparse and the available solution paradigms relatively homogeneous. Consequently, noisy or insufficient supervision may lead models to adopt inefficient search strategies, generate unsupported reasoning (i.e., hallucination), or produce answers prematurely before adequate evidence has been gathered.

A growing body of recent research \cite{team2026mirothinker, chu2026redsearcher, deng2026fort, du2026openseeker, zhu2026marco, hu2025step} indicates that synthetic task generation and automated verification can serve as scalable alternatives to human annotation. 
Nevertheless, existing approaches continue to face several important limitations. 
First, many synthetic data pipelines emphasize final-answer correctness while paying insufficient attention to the quality of intermediate search decisions and transitions between pieces of evidence.
Second, weak filtering procedures often retain shallow, repetitive, or redundant trajectories, thereby providing limited supervision for learning effective exploration strategies.
Third, existing training corpora frequently underrepresent challenging, long-horizon tasks that require sustained information integration, iterative reasoning, and adaptive retrieval planning.

In this technical report, we introduce \textbf{SearchArt}, a scalable framework for training long-horizon search agents through verification-driven synthetic task generation and a multi-stage post-training pipeline.
SearchArt constructs large-scale datasets for complex \textbf{search}-, \textbf{research}- and \textbf{user}-oriented tasks by synthesizing diverse information-seeking QA pairs, together with their corresponding search trajectories, from web documents and automatically generated evidence graphs.
To ensure the quality of the synthesized data, we develop a verification pipeline that jointly evaluates QA consistency, trajectory reliability, and the relevance of retrieved evidence. The verified trajectories are subsequently utilized for supervised fine-tuning and reinforcement learning–based policy optimization.
Through training process, the agent learns to coordinate retrieval and reasoning over long-horizon tasks, selectively expand evidence branches, and adaptively allocate search budgets according to task complexity. 
In contrast to prior approaches that primarily optimize for final-answer supervision, SearchArt explicitly models and optimizes the quality of intermediate search and reasoning behaviors. This design yields more stable, efficient, and reliable search policies for complex information-seeking tasks.

Building on this framework and Qwen3.5-27B model, we evaluate our approach on a diverse set of open-domain deep-research and multi-hop reasoning benchmarks. 
Experimental results show that SearchArt, despite having only 27B parameters, achieves strong performance across multiple benchmarks, scoring \textbf{74.39 on BrowseComp-ZH}, \textbf{70.06 on BrowseComp}, \textbf{63.49 on BrowseComp-Plus}, \textbf{64.0 on WideSearch}, and \textbf{52.55 on DeepResearch-Bench}.
These findings demonstrate that SearchArt matches or even surpasses frontier closed-source agents on both deep search and deep research tasks, and significantly improves task-completion accuracy relative to existing search agents of comparable scale.
Overall, our work highlights the importance of scalable synthetic supervision and verification-centric training in developing the next generation of deep-research agents. We believe these findings provide a practical foundation for building more capable, efficient, and trustworthy search systems that can operate autonomously over complex information environments.

\section{Scalable Long-Horizon Task Synthesis \& Verification}
In this section, we present a scalable framework for synthesizing long-horizon tasks for training and evaluating search agents. The framework addresses three objectives: generating realistic information-seeking tasks grounded in open-world knowledge, constructing reasoning and retrieval trajectories with controllable complexity, and ensuring answer faithfulness and reliable verification throughout synthesis. To this end, we develop three complementary QA synthesis pipelines: (1) a DeepSearch-oriented pipeline for scalable generation of retrieval-intensive questions with increasing complexity; (2) a DeepResearch-oriented pipeline for research-style tasks requiring extended exploration, multi-source evidence aggregation, and long-context reasoning; and (3) a Real-world User Experience–oriented pipeline for difficult questions that preserve natural user intent, practical decision needs, and realistic interaction patterns while requiring substantial search depth and breadth.


\subsection{DeepSearch QA Synthesis}
\begin{figure}[!t]
    \centering
    \includegraphics[width=1.0\linewidth]{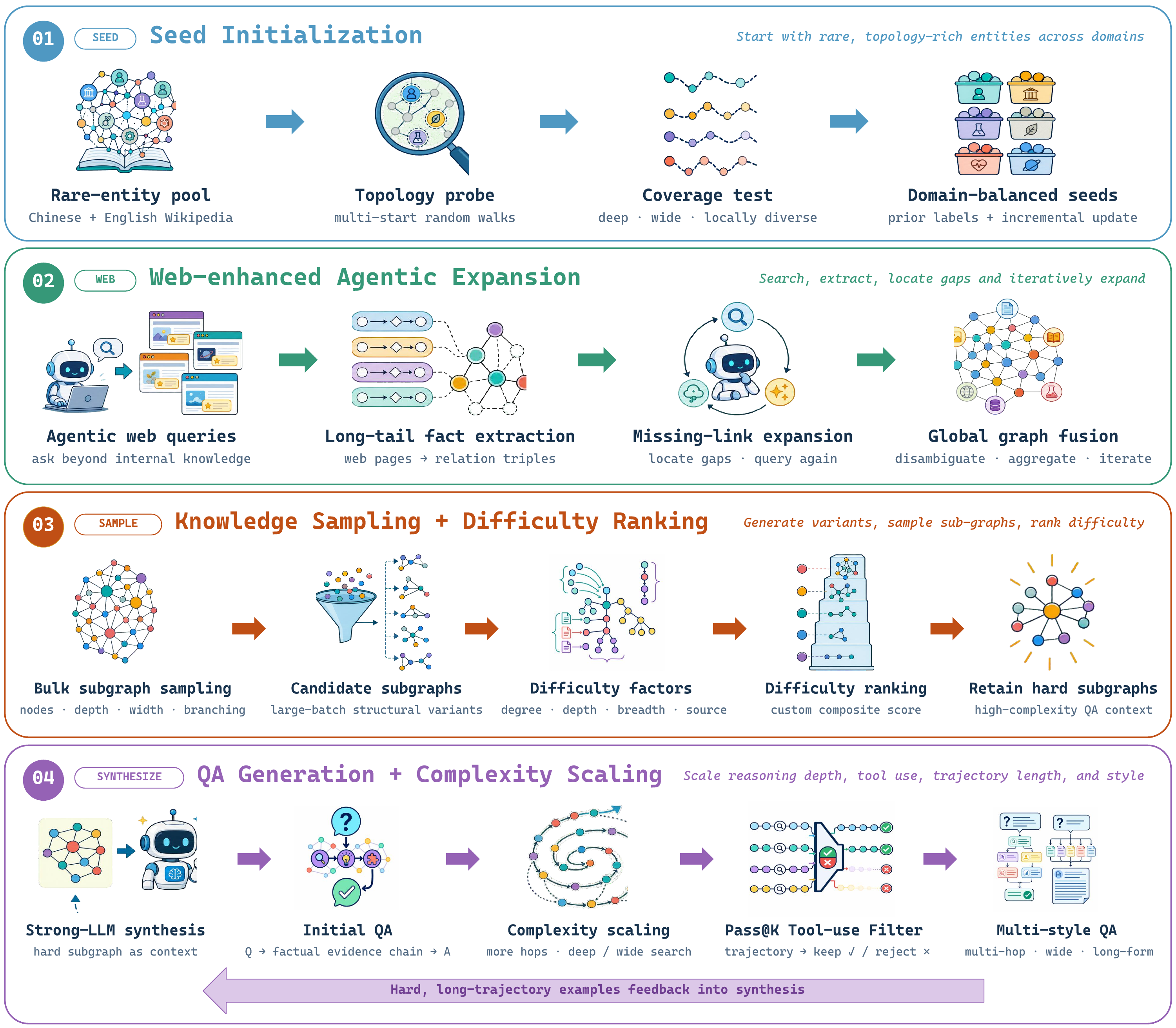}
    \caption{Overview of the DeepSearch QA Synthesis.}
    \label{fig:deepsearch_qa_synthesis_framework}
\end{figure}

We propose a shortcut-resistant and evidence-dispersed synthesis framework for constructing challenging long-horizon DeepSearch QA tasks. The framework is designed to substantially reduce reliance on parametric knowledge, surface-level retrieval cues, or isolated evidence snippets, while encouraging sustained web exploration, dispersed evidence aggregation, and compositional reasoning. As illustrated in Figures~\redref{fig:deepsearch_qa_synthesis_framework} 
and~\redref{fig:synthetic_task_qa_trajectory_distribution_bright}, the pipeline follows a progressive synthesis process that transforms selected seed entities into executable long-horizon search tasks. It first initializes seeds that balance knowledge novelty, structural connectivity and domain coverage, and then expands them into a large-scale evidence-rich knowledge graph through iterative web-enhanced agentic exploration. Structurally complex subgraphs with dispersed evidence are then sampled and ranked by their estimated reasoning and retrieval difficulty. Finally, these knowledge regions are converted into QA instances through a complexity-scaling process that increases reasoning-chain length, evidence dispersion, and search-trajectory difficulty. This process yields diverse and controllably difficult DeepSearch QA data for faithfully evaluating agent capabilities in open-world information acquisition, evidence integration, and multi-step reasoning.
We give QA examples in Appendix \redref{appendix:DeepSearch QA}.

\subsubsection{Seed Initialization}
Seed initialization determines the distribution from which long-horizon search tasks are drawn, and thus critically shapes their difficulty, diversity, and scalability. Instead of sampling at random, we deliberately curate a pool of \emph{long-tail} entities selected along three complementary axes—limited parametric exposure, structural richness, and balanced domain coverage—which we detail below. These entities then seed the subsequent web-enhanced agentic knowledge expansion stage. 

\paragraph{Limited Exposure in Parametric Knowledge.}
Entities that are extensively represented in a model's parametric knowledge may permit shortcut solutions, allowing the model to answer directly from memorization rather than genuine external retrieval. To suppress such shortcuts, we sample long-tail entities from large-scale multilingual knowledge graphs constructed from Wikipedia, prioritizing those with low in-degree and out-degree. Grounding tasks in these less prominent entities substantially increases the likelihood that successful completion requires actively acquiring and integrating external evidence, thereby preserving the retrieval-centric nature of the benchmark.

\paragraph{Sufficient Structural Richness.} 
Rare entities often reside in structurally sparse regions of the knowledge graph, where limited relational context constrains the construction of multi-hop reasoning chains and complex evidence dependencies. To quantify the context-expansion potential of each candidate, we perform multiple randomized graph walks originating from the entity and analyze the topology of the resulting local neighborhoods. Seeds that consistently reach structurally rich subgraphs—characterized by diverse relation types, branching evidence paths, and extended multi-hop connectivity—are assigned higher priority, ensuring that each seed can anchor a sufficiently deep and interconnected reasoning task.

\paragraph{Balanced Domain Coverage.}
Beyond structural complexity, broad topical coverage is essential for scalable task synthesis. Because graph-based sampling tends to concentrate on densely connected domains while overlooking specialized areas, we organize seed entities within a hierarchical domain taxonomy that couples predefined categories with regions discovered dynamically during expansion. Domain-specific sampling quotas are then adjusted on the fly to maintain balanced representation, preventing the resulting task set from collapsing onto a handful of popular topics and instead promoting even coverage across the knowledge landscape.

\subsubsection{Web-enhanced Agentic Knowledge Expansion}

\begin{figure}[!t]
    \centering
    \includegraphics[width=1\linewidth]{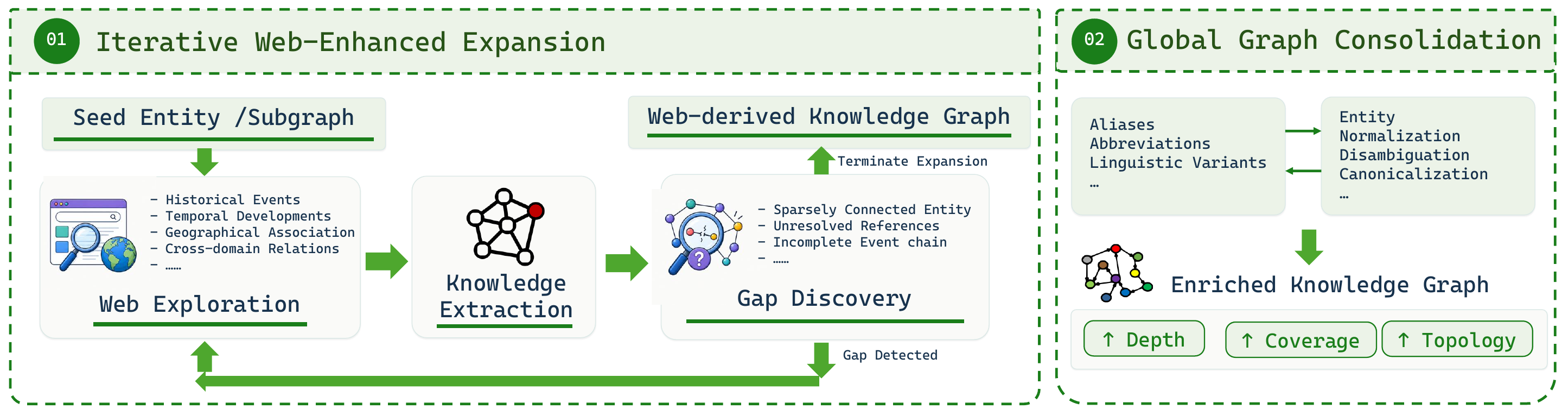}
    \caption{Procedure of web-enhanced knowledge expansion.}
    \label{fig:placeholder}
\end{figure}

Synthesizing data for search agents requires both broad exploration and deep multi-hop reasoning, demanding comprehensive open-world knowledge coverage and diverse relational connectivity. Existing approaches typically rely on graph traversal over pre-constructed knowledge graphs, restricting exploration to predefined relations and a static knowledge coverage \cite{qwenbai2023qwen,chu2026redsearcher}.
To address these limitations, we develop an explorer agent that autonomously acquires multi-source information and progressively expands a large-scale knowledge graph, which subsequently serves as the basis for the graph sampling stage. 
We give the overview of Web-enhanced Knowledge Expansion in Figure \redref{fig:placeholder}.

Starting from a seed entity or an intermediate knowledge subgraph, the explorer agent iteratively expands the graph through a cycle of query generation, web retrieval, knowledge extraction, graph construction, and knowledge-gap discovery. At each iteration, the agent generates queries that go beyond the information already encoded in the current graph. Instead of focusing only on direct attributes of the target entity, it explores broader informative contexts, such as related entities, historical events, organizational affiliations, temporal developments, geographical associations, and cross-domain relations. This iterative process enables the framework to acquire knowledge that is difficult to obtain through local graph traversal alone. Newly extracted entities and relations are continuously integrated into the evolving graph, enabling the search frontier to move beyond local neighborhoods and form richer topologies such as multi-hop chains, branching paths, convergent evidence trails, and densely connected subgraphs.

\paragraph{Web-guided Knowledge Acquisition.} For each generated query, the agent retrieves evidence from external web pages and extracts candidate factual statements from the returned content. To maximize the informational value of acquired information, the extraction process prioritizes uncommon facts and non-trivial entity relations over frequently repeated surface-level descriptions. Each extracted statement is converted into a structured relation representation consisting of: (1) Source entities, (2) Target entities, (3) Relation types, (4) Textual descriptions, and (5) Supporting evidence references. Relations extracted from each retrieval source are subsequently organized into source-specific local knowledge graphs.

\paragraph{Knowledge-gap Discovery \& Targeted Expansion.} A major challenge is that Evidence retrieved from the open web often yields incomplete local graph structures. For example, an entity may be introduced without sufficient contextual connections, an event sequence may contain missing intermediate steps, or a relationship may be supported only from one direction. To address this issue, we introduce a knowledge-gap discovery mechanism that analyzes the current graph and identifies regions with potentially requiring further exploration. Such regions include sparsely connected entities, unresolved references, incomplete event chains, and partially observed relationship structures. Based on these signals, the agent formulates targeted follow-up queries and initiates additional retrieval rounds to acquire complementary evidence. This mechanism allows the expansion process to progress beyond immediate entity neighborhoods and gradually reveal broader, more coherent evidence networks.

\paragraph{Global Graph Consolidation.} After multiple expansion rounds, knowledge graphs derived from heterogeneous web sources is integrated into a unified global knowledge graph. Since the same real-world entity may appear under different aliases, abbreviations, or linguistic variants, we perform entity normalization, disambiguation, and canonicalization before graph integration. Duplicate or semantically equivalent relations are aggregated while preserving source-level provenance and supporting evidence. The final graph therefore combines evidence acquired across numerous search trajectories and domains, yielding a substantially larger and structurally richer knowledge space than the initial seed graph.

The consolidated global graph provides three properties that are particularly important for long-horizon task synthesis: sufficient depth to support extended multi-step reasoning, broad coverage to enable diverse search trajectories, and heterogeneous local topologies to facilitate the construction of complex search objectives. These properties form the basis of the subsequent knowledge sampling and difficulty-ranking stage.

\subsubsection{Knowledge Sampling with Difficulty Ranking}
After constructing the global knowledge graph, the next challenge is to identify substructures that can yield high-quality, long-horizon search tasks. Although the expansion stage produces a large and heterogeneous graph, many local regions are unsuitable for complex QA synthesis. Some subgraphs contain only shallow relational patterns, whereas others, despite their apparent structural complexity, can be resolved using a small number of highly concentrated evidence sources. We therefore introduce a large-scale subgraph sampling and difficulty-ranking procedure that preferentially retains knowledge regions exhibiting both complex relational structures and dispersed evidence provenance.

Rather than generating questions directly from the entire graph, we first sample extensive subgraphs under configurable structural constraints. Starting from randomly selected root entities, the sampler iteratively expands neighboring nodes while controlling the maximum number of entities, expansion depth, branching width, and node-selection policy at each step. This process yields a large collection of candidate subgraphs with diverse topological structures and evidence distributions. These subgraphs serve as intermediate reasoning templates from which complex search tasks are subsequently synthesized.

To identify the most challenging candidates, we evaluate the difficulty of each sampled subgraph along two complementary dimensions: structural complexity and evidence dispersion. The former characterizes the complexity of the underlying reasoning structure, whereas the latter estimates the extent to which the supporting evidence must be acquired through multiple independent search paths during inference.

\paragraph{Structural Complexity.} 
The first component quantifies the topological complexity of the sampled subgraph. Intuitively, reasoning becomes more challenging when entities are embedded in densely interconnected relational structures, and when deriving an answer requires integrating information across deeper and broader dependency paths. 
To capture this difficulty, we characterize the structural complexity of the underlying reasoning graph along three complementary dimensions: connectivity, depth, and width, which respectively reflect relational density, the length of dependency chains, and the breadth of parallel evidence branches.
\begin{itemize}[leftmargin=1em, listparindent=\parindent]
\item \textbf{Average Node Connectivity.} For a sampled subgraph $G=(V,E)$, we compute the average node degree:
\begin{equation}
C_{\text{degree}}(G)=\frac{1}{|V|}\sum_{v\in V}\deg(v)
\end{equation}
Higher average connectivity generally indicates denser relational coupling, richer local topology, and a greater diversity of potential reasoning paths. Such structures are more likely to exhibit intersecting evidence chains, alternative traversal routes, and complex dependencies among relational paths, thereby increasing the difficulty of multi-step reasoning.

\item \textbf{Depth Complexity.} We further measure the minimum expansion depth required to reach all sampled entities from the root of the subgraph. A greater depth indicates longer chains of evidence acquisition, suggesting that the final task is more likely to require iterative, multi-step information discovery rather than localized retrieval.

\item \textbf{Width Complexity.} While depth reflects the length of the reasoning chain, width characterizes its breadth. Subgraphs with greater width contain more parallel evidence branches that may need to be explored, compared, and reconciled. Accordingly, a larger minimum width expands the combinatorial space of candidate reasoning trajectories, and reduces the likelihood that a question can be solved through a narrow shortcut.
\end{itemize}

\paragraph{Evidence Dispersion.}
Structural complexity alone does not necessarily translate into search difficulty. In open-web settings, even a logically complex reasoning structure may remain relatively easy to solve when most of the supporting evidence can be retrieved from only a few search results or source documents. To capture this complementary dimension of difficulty, we introduce \textbf{Evidence Source Dispersion (ESD)}, which quantifies the extent to which the evidence required for solving a question is distributed across heterogeneous information sources.

Each entity and relation in the global knowledge graph is associated with provenance metadata, including the identifier of the search query through which it was discovered and the identifier of its source document. For each sampled subgraph, we aggregate these provenance assignments into source groups and examine the resulting evidence distribution. ESD is then computed using the coefficient of variation of the evidence counts across source groups, thereby characterizing the degree of concentration or dispersion of supporting evidence across queries and documents.

Let $n_i$ denote the number of facts originating from source group $i$. We define:
\begin{equation}
\text{ESD}(G)=\frac{\sigma(n)}{\mu(n)}
\end{equation}
where $\sigma(n)$ and $\mu(n)$ are the standard deviation and mean of the source-frequency distribution, respectively.

Subgraphs with higher evidence dispersion typically require information gathering from multiple independent searches and source documents, rather than from a single highly informative page. These instances are more consistent with realistic deep-search scenarios, in which the process of locating, validating, and synthesizing distributed evidence constitutes a substantial component of the overall task difficulty.

\paragraph{Difficulty Ranking \& Selection.}
The final difficulty score integrates structural complexity and evidence dispersion into a unified ranking criterion. Candidate subgraphs are ranked accordingly, and only the highest-scoring instances are retained for downstream QA synthesis. This selection procedure enriches the generated dataset with long-horizon reasoning structures, multi-source evidence dependencies, and topologically diverse search trajectories.

The retained subgraphs consequently provide a high-quality basis for QA generation, enabling the synthesis pipeline to construct challenging search tasks that require both complex multi-step reasoning and sustained evidence acquisition across heterogeneous sources.

\subsubsection{QA generation with Complexity Scaling}
Given the retained high-difficulty subgraphs, the final stage converts structured knowledge regions into executable DeepSearch QA instances. Rather than merely generating questions that involve multiple facts, we aim to synthesize tasks whose solutions require sustained evidence acquisition, multi-step reasoning, and the resolution of underspecified or obfuscated clues. We therefore introduce a complexity-scaling QA generation procedure that progressively increases task difficulty through subgraph-grounded synthesis, iterative reasoning-chain expansion, and hard-trajectory conditioning.

\paragraph{Subgraph-grounded QA generation.} For each selected subgraph, we prompt a strong LLM to synthesize multiple candidate instances, each comprising a question, an answer, and a factual reasoning chain. The sampled subgraph serves as the factual backbone of the instance, while the strong LLM converts its explicit entities and relations into natural-language search objectives. To prevent direct entity matching and shallow retrieval, the generator is instructed to obfuscate surface-level constants, including entity names, relation descriptions, dates, locations, and numerical values where appropriate. Consequently, the resulting questions typically cannot be solved by directly searching for the answer entity or a single explicit clue. Instead, the solver must identify intermediate targets from indirect descriptions, relational constraints, and partially specified evidence, often through inverse or abductive reasoning.

\paragraph{Factual Reasoning-chain Construction.} Each generated instance is accompanied by a factual reasoning chain grounded in the underlying subgraph. The chain decomposes the solution process into a sequence of evidence-dependent steps, each corresponding to a retrieval operation, clue-resolution step, or inference over retrieved facts. We favor longer chains when they remain factually valid, logically necessary, and free from artificial ambiguity. This design preserves verifiability while requiring the agent to perform long-horizon search and reasoning rather than relying on isolated fact retrieval.

\paragraph{Iterative Complexity Expansion.} Initial generations may still admit retrieval shortcuts or contain reasoning paths of insufficient depth. We therefore apply an iterative enhancement procedure. Given the sampled subgraph and the initially generated QA instances, a strong LLM is prompted to extend, refine, or recombine their factual reasoning chains by incorporating additional entities, relations, and evidence dependencies from the same subgraph. This process transforms relatively direct questions into more demanding search tasks involving broader exploration, deeper dependency chains, and more careful evidence aggregation. An expanded instance is retained only when its answer remains uniquely determined by the supporting evidence and every added reasoning step is factually grounded and relevant to the final solution.

\paragraph{Hard-trajectory-conditioned Generation.} To further increase the lower bound of task difficulty, we incorporate previously synthesized hard examples into the generation process. For QA instances produced in earlier rounds, we collect long-horizon solving trajectories from strong LLM agents and estimate task difficulty using Pass@$K$-based stratification. QA Instances that remain difficult under multiple sampled solving attempts are treated as high-difficulty examples. Their long-horizon trajectories are then sampled as in-context demonstrations for subsequent rounds of QA generation. These demonstrations expose the generator to empirically difficult search patterns, including delayed clue resolution, multi-branch exploration, ambiguous intermediate entities, repeated query reformulation, and the reconciliation of evidence distributed across distant sources. Hard-trajectory conditioning therefore shifts the generation distribution toward instances that are difficult not only according to static graph properties, but also according to observed agent behavior. Consequently, the synthesized tasks place greater demands on long-horizon planning, adaptive search, intermediate-state tracking, and comprehensive evidence integration.

\paragraph{Multi-style QA Synthesis.} In addition to increasing reasoning complexity, we diversify the linguistic form, interaction style, and expected answer format of the synthesized instances. If all questions follow a benchmark-style multi-hop template, a post-trained agent may overfit to a narrow distribution of clue-dense and explicitly compositional inputs. We therefore prompt the generator to synthesize QA instances in multiple styles. Besides complex multi-hop DeepSearch questions, it generates natural and conversational queries whose information needs are less formally specified. Such questions more closely resemble real user queries, in which the intent may be implicit, relevant constraints may be scattered across the prompt, and the solver must determine which evidence is necessary before constructing an answer.

We further diversify the expected response format. In addition to short factual answers, the generator produces paragraph-level answers that require broader evidence aggregation and explanation. These instances may depend less on a single deep reasoning chain and more on wide search ability: the agent must identify multiple relevant sources, reconcile potentially dispersed evidence, and organize the findings into a coherent, evidence-grounded answer. This design complements depth-oriented multi-hop reasoning with breadth-oriented information synthesis. Collectively, the generated data cover a wider range of practical DeepSearch scenarios, including precise fact tracing, ambiguous intent interpretation, broad information gathering, and evidence-grounded answer synthesis.

The final output of this stage is a collection of $\langle Question, Answer, Factual \ Reasoning \ Chain \rangle$ triples with progressively scaled complexity. These instances are grounded in high-difficulty subgraphs, strengthened through iterative reasoning-chain expansion, and further calibrated by hard trajectory contexts. They provide challenging supervision signals for training and evaluating deep search agents.

\subsection{DeepResearch QA Synthesis}
In addition to graph-grounded DeepSearch QA synthesis, we construct a complementary collection of DeepResearch QA pairs from high-quality, long-form research documents. Unlike fact-centric search questions are typically grounded in localized knowledge subgraphs, DeepResearch questions require agents to explore a broader body of literature, reconcile heterogeneous evidence, and produce comprehensive answers comparable to an expert-written survey or or analytical reports. We therefore regard existing high-quality surveys, review articles, research reports, and analytical papers as naturally occurring reference answers for deep research tasks.

\begin{figure}[!t]
    \centering
    \includegraphics[width=1.0\linewidth]{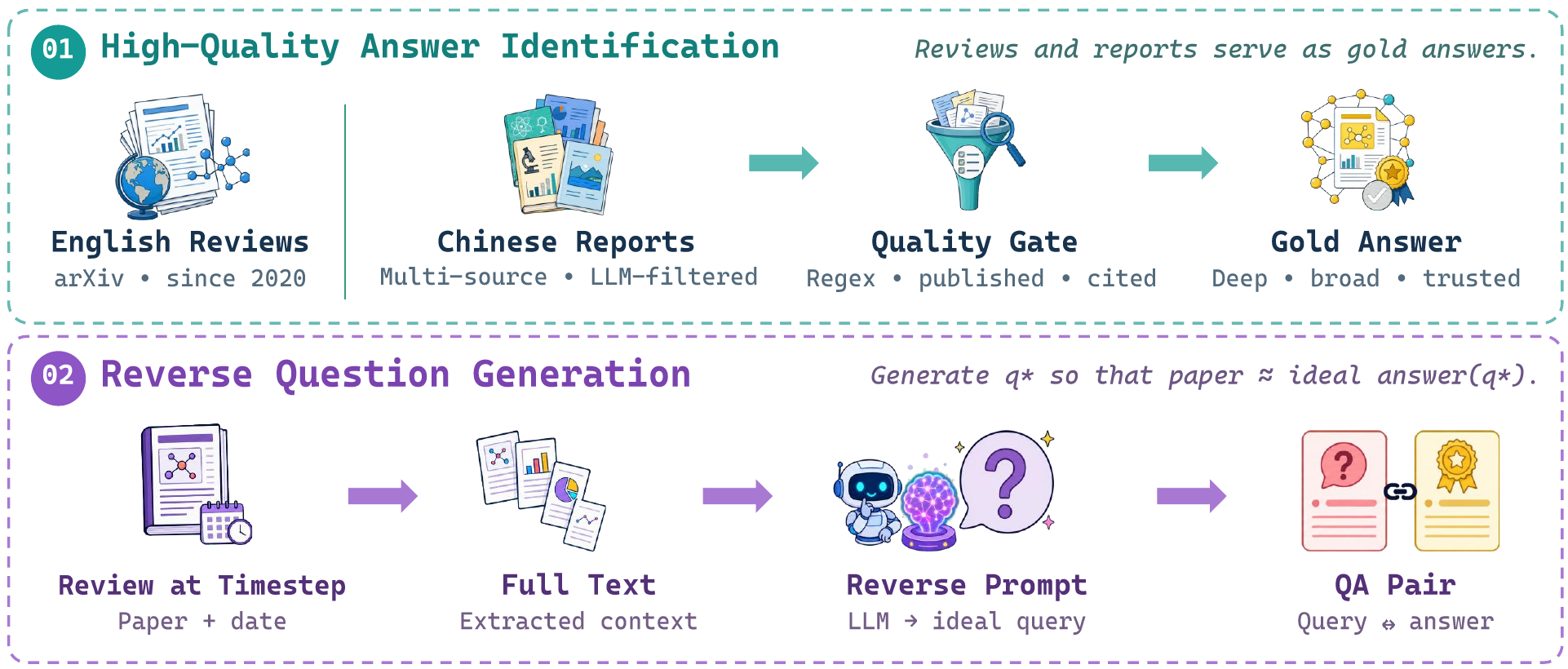}
    \caption{Overview of the DeepResearch QA Synthesis.}
    \label{fig:deepresearch_qa_synthesis_framework}
\end{figure}

The synthesis pipeline comprises two stages: high-quality answer identification and reverse question generation. In the first stage, we collect long-form documents that provide systematic and well-supported syntheses of specific research topics. In the second stage, we apply reverse prompt engineering to infer the underlying research question for which each document would constitute an ideal answer. This process transforms existing expert-authored documents into supervised question–answer pairs, providing valuable training and evaluation data for deep research agents.
We give the overview and final topic distribution of the DeepResearch QA synthesis in Figure \redref{fig:deepresearch_qa_synthesis_framework} and \redref{fig:deepresearch_theme_distribution}. We give QA examples in Appendix \redref{appendix:DeepResearch QA}

\subsubsection{High-quality Answer Identification}
The first stage identifies high-quality long-form documents that can serve as reference answers. We focus on surveys, review articles, research reports, and analytical studies because these documents typically synthesize extensive evidence, compare multiple lines of research, and organize information around a clearly defined topic. Unlike short factual answers, such documents more closely reflect the capabilities expected of DeepResearch agents, including sustained information seeking, multi-source integration, critical analysis, and structured synthesis.

For English-language documents, we begin with a comprehensive snapshot of arXiv metadata and retain papers submitted from 2020 onward. Candidate survey-style papers are identified by applying regular-expression filters to their titles, using indicative terms such as "survey", "review", and related expressions. Because title-based retrieval may produce false positives, we subsequently apply additional quality filters. Specifically, when such information is available, we prioritize papers with evidence of peer-reviewed publication, formal acceptance, or substantial citation impact. These signals help distinguish externally validated and influential studies from manuscripts that merely exhibit a survey-like format.

For Chinese-language documents, we collect a broader range of in-depth reports, surveys, technical analyses, and domain-specific research documents from multiple publicly accessible sources. As these documents are distributed across heterogeneous platforms and often lack standardized metadata, conventional metadata-based filtering is insufficient. We therefore employ LLMs to conduct content-level quality filtering. Each document is evaluated in terms of analytical depth, structural coherence, factual density, topical coverage, and the quality of its supporting evidence. Documents that are superficial, predominantly promotional, weakly substantiated, poorly structured, or of limited analytical value are excluded.

Through this process, we obtain a multilingual pool of high-quality long-form answers. These answers cover both academic and applied domains, and provide naturally challenging targets for deep research question generation.

\subsubsection{Reverse Question Generation}
A survey, review, or research report can be regarded as a comprehensive answer to a deep research question under a specific temporal knowledge boundary. Motivated by this perspective, we adopt a reverse question-generation strategy to construct research queries from collected reference documents. Given the publication date and parsed full text of a document, a strong LLM is prompted to infer a research query for which the document would constitute a high-quality reference answer.

Specifically, the model is prompted to formulate a query that reflects the document's central research objective, thematic scope, and expected level of analytical depth. The generated query should not simply request a summary of the source document. Instead, it should define an open-ended research task that naturally requires extensive information retrieval, synthesis, and organization. For example, a survey on retrieval-augmented generation may be transformed into a query requesting a systematic analysis of its major methodological paradigms, evaluation benchmarks, practical limitations, and future research directions.

To preserve temporal consistency between the generated question and its reference answer, the query is conditioned on the publication date or latest update date of the source document. This date serves as the knowledge cutoff for the corresponding research task, ensuring that the expected answer is evaluated against the information available when the reference document was produced. When necessary, the prompt explicitly requires that the scope of the answer be restricted to developments reported before this cutoff date. This design prevents subsequent findings from being incorrectly incorporated into the reference answer and reduces temporal mismatch between questions and answers.

We further introduce constraints to prevent the generated query from revealing the identity of the source document. In particular, the question must not include the exact title, author names, arXiv identifier, report name, or other distinctive surface-level metadata that would enable a solver to retrieve the reference answer directly. Instead, it should express the underlying research objective in natural language. This constraint discourages document-retrieval shortcuts and creates more realistic deep research tasks, in which an agent must independently search for relevant evidence, reconcile information from multiple sources, and synthesize a coherent response.

This stage ultimately produces a collection of document-grounded QA pairs. Each answer corresponds to an existing high-quality survey, review, research report, or analytical document, whereas each question is a reverse-generated deep research query aligned with the document's content and temporal scope. These pairs provide complementary supervision to graph-grounded DeepSearch QA. Graph-based synthesis primarily supports controllable, fine-grained factual reasoning over structured evidence chains, whereas document-based DeepResearch synthesis emphasizes broad topical coverage, long-form evidence integration, and expert-level organization of complex information.

\begin{figure}[!t]
    \centering
    \includegraphics[width=1.0\linewidth]{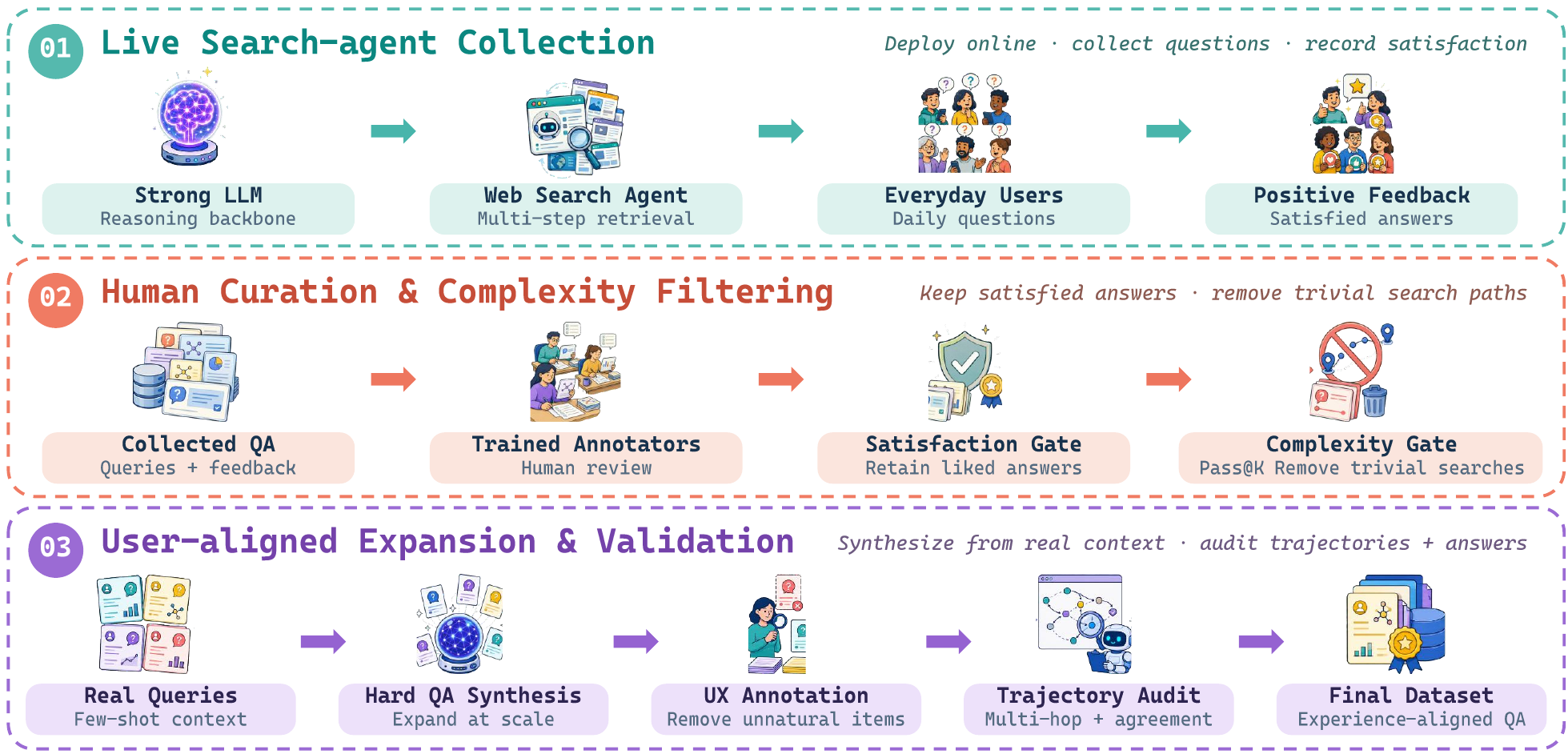}
    \caption{Overview of the Real-world User Experience–oriented QA Synthesis.}
    \label{fig:Real_world_User_Experience_QA_synthesis_framework}
\end{figure}

\subsection{Real-world User Experience–oriented QA}
Although synthetic long-horizon search tasks enable precise control over question difficulty and reasoning structure, they do not fully reflect how users interact with search agents in real-world scenarios. Existing search-agent data synthesis pipelines primarily focus on benchmark-style questions, such as multi-hop reasoning tasks constructed with intentionally obfuscated clues. While such questions are valuable for evaluating search depth and resistance to reasoning shortcuts, they can differ substantially from daily information-seeking needs. Real users often pose open-ended, preference-sensitive, and decision-oriented questions that require broad evidence collection, contextual understanding, and grounded synthesis.

To narrow this gap, we construct a complementary collection of Real-world User Experience-oriented QA data. Our objective is to develop challenging questions that preserve natural user intent, practical decision-making needs, and realistic interaction patterns, while still requiring substantial search depth and breadth. These questions often involve scenarios such as evaluating recently released products, conducting technical comparisons, planning travel, making financial decisions, interpreting policy changes, developing learning strategies, and balancing personalized trade-offs.
For example, a user may ask whether replacing two 27-inch 4K monitors with a single 32-inch 4K monitor would be preferable for programming, spreadsheet work, and occasional gaming, while considering products released or expected around May 2026, cost-effectiveness, workspace constraints, and real-world user feedback. Answering such a question cannot be reduced to a single factual lookup. Instead, it requires temporal awareness, multi-source evidence retrieval, structured comparison, and user-centered reasoning. By incorporating these realistic and decision-intensive queries, this dataset complements synthetic benchmark-style tasks and provides a more direct basis for improving the practical utility of search agents.
We give the overview of the Real-world User Experience–oriented QA synthesis in Figure \redref{fig:Real_world_User_Experience_QA_synthesis_framework}.
We give QA examples in Appendix \redref{appendix:Real-world User Experience–oriented QA}

\subsubsection{User-facing Search Agent Deployment}
We deploy the search-agent inference framework through an online web interface powered by a strong LLM. Rather than interacting with artificially constructed benchmark prompts, users are encouraged to submit natural, daily information-seeking questions. For each query, users can inspect the generated answer and provide feedback on its quality, usefulness, factual accuracy, and overall satisfaction.

This deployment enables the collection of queries that more closely reflect real-world information needs and interaction patterns. Compared with manually designed benchmark tasks, user-generated queries exhibit greater diversity in intent, less standardized phrasing, and more realistic constraints. They often involve underspecified preferences, implicit assumptions, time-sensitive information, multi-objective comparisons, and subjective assessments of utility. These characteristics provide valuable training signals for developing search agents that can generalize beyond rigid benchmark settings and respond effectively to heterogeneous user needs.

\subsubsection{Human Filtering and Quality Control}
After collecting a sufficiently large pool of user queries and associated feedback annotations, we conduct a human-in-the-loop filtering process to construct a high-quality seed set. Multiple trained annotators independently review the collected examples and retain queries for which users indicate satisfaction with the quality and usefulness of the corresponding answers. We exclude examples that are trivial, highly subjective and lack verifiable evidence, or can be resolved through only one or two superficial search steps.

The filtering process is guided by two primary criteria. First, each query should reflect a realistic information need and be formulated in a manner consistent with how users naturally express such needs in practice. Second, answering the query should require substantive search and reasoning, such as retrieving up-to-date information, consulting and comparing multiple sources, synthesizing evidence across documents, or reconciling conflicting claims. This procedure ensures that the retained examples are both representative of real-world user requests and sufficiently search-intensive.

\subsubsection{User-style QA Synthesis}
Although directly collected user queries offer high authenticity, they remain limited in both scale and domain coverage. To broaden the dataset while preserving realistic user intent, we use the collected queries as in-context demonstrations for further synthesis. Specifically, we integrate these examples into the DeepSearch QA Synthesis pipeline described above and prompt a strong LLM to generate additional questions that retain the linguistic style, intent structure, and practical orientation of real user requests, while incorporating the difficulty controls introduced by graph-grounded and search-intensive QA synthesis.

This procedure produces questions that are challenging not only in terms of reasoning complexity, but also from the perspective of real-world information seeking. The synthesized queries commonly involve constrained decision-making, multi-source comparison, recent or evolving information, and nuanced trade-offs between competing objectives. Unlike benchmark-style questions that rely on artificial obfuscation to increase difficulty, these examples are designed to resemble authentic user requests for which accurate and comprehensive answers require substantial search, evidence integration, and contextual reasoning.

\subsubsection{Post-generation Validation}
To ensure data quality, the synthesized user-oriented questions undergo a multi-stage validation procedure. First, human annotators remove questions that do not reflect natural interaction patterns, impose implausible or contrived constraints, or otherwise appear overly artificial. Second, a strong LLM is used to generate search trajectories and candidate answers for each retained question. We discard examples whose solution requires only a shallow search process, whose supporting evidence can be retrieved through a single straightforward query, or whose generated answers are inconsistent across repeated attempts.

This filtering and validation procedure yields a collection of real-world, user-experience-oriented QA examples characterized by three key properties: natural user intent, practical answer utility, and substantial search complexity. The resulting dataset complements the synthetic DeepSearch and document-grounded DeepResearch QA data by grounding search-agent training in realistic usage scenarios. Consequently, the trained agent is encouraged not only to solve formally challenging tasks, but also to provide useful, evidence-grounded answers to the complex and practical questions encountered in real-world interactions.

\subsection{Multi-Stage Task Verification}
The synthetic QA construction process is designed to increase the difficulty of search and reasoning through mechanisms such as clue obfuscation, multi-hop dependency construction, and evidence integration across multiple knowledge sources. Although these mechanisms are effective in producing challenging tasks, they may also introduce noise and reduce data reliability. For example, some generated QA instances may remain solvable without long-horizon search, whereas others may be ill-posed because the question does not uniquely determine the intended answer. In addition, the target answer may be unverifiable, ambiguous, or inconsistent with the available evidence. To ensure that the resulting dataset is both challenging and reliably verifiable, we introduce a multi-stage task verification pipeline.

The pipeline comprises five stages: rule-based and model-based QA cleaning, direct-reasoning verification using a baseline model, agentic tool-use verification using the same baseline model, agentic verification and trajectory synthesis using a strong model, and final trajectory selection. Together, these stages progressively remove invalid or low-quality instances, estimate task difficulty, verify QA consistency and evidence support, and retain high-quality search trajectories for subsequent training.

\subsubsection{QA Cleaning}
We first apply an automated cleaning procedure to the generated QA pairs to remove malformed and low-quality instances. This stage evaluates surface-level quality, language consistency, length constraints, and content validity. Each QA pair must be written in one of the supported languages, such as Chinese or English; instances containing unsupported languages or inappropriate language mixing are discarded. To prevent questions from becoming overly fragmented or unnecessarily complex, each question is restricted to no more than two interrogative sentences. We also impose a maximum answer length, such as 1,000 tokens, to maintain concise reference answers and prevent them from expanding into uncontrolled long-form documents.

We further remove QA pairs that reveal the synthesis process or rely on unavailable construction context. For example, instances containing expressions such as "according to the given text", "in the knowledge graph", or "based on the above information" are excluded, as these template artifacts indicate that the question is not fully self-contained and depends on hidden contextual information. We additionally discard answers that constitute refusals, non-answers, or statements that the question cannot be answered. Together, these filtering criteria ensure that the retained QA pairs are linguistically well formed, self-contained, answerable, and suitable for search-based evaluation.

\subsubsection{Stage-I Verification: Baseline Direct Reasoning}
The first verification stage estimates whether a task can be solved using parametric knowledge alone. For each QA pair, a baseline LLM is prompted to answer the question directly without access to external search tools. To account for stochastic variation in model generation, we independently sample $N$ responses. Each response is then evaluated against the reference answer using an automated answer-matching verifier.

An instance is retained for subsequent verification only if the baseline model fails to produce a correct answer in all $N$ attempts. Such instances are less likely to be solvable through internal knowledge and are therefore more likely to require external information retrieval and multi-step reasoning. By contrast, instances answered correctly in at least one direct-reasoning attempt may contain overly revealing clues, depend primarily on common knowledge, or otherwise lack sufficient search difficulty; these instances are removed or assigned lower priority.

\subsubsection{Stage-II Verification: Baseline Agentic Tool Use}
Failure under direct generation alone does not establish that a task is both valid and well-posed. A question may remain unsolved because it is ambiguous, lacks sufficient supporting evidence, admits multiple plausible answers, or is inconsistent with its reference answer. To distinguish genuinely challenging tasks from invalid ones, we next allow the same baseline model to solve each question using agentic tools. In each of $N$ independent attempts, the model may autonomously invoke search tools, producing complete solution trajectories and final answers.

Let $n$ $\in$ $\{0,1,\ldots,N\}$ denote the number of attempts whose final answers match the reference answer. When $n>0$, at least one valid solution path has been identified under tool-use reasoning. This provides evidence that the question and reference answer are aligned and that the answer can be recovered from external evidence. By contrast, when $n=0$, the instance may either be too difficult for the baseline agent or be invalid because of reference-answer inconsistency, insufficient evidence, ambiguity, or non-unique solutions. Such cases are therefore subjected to stronger verification in the subsequent stage.

The empirical success rate $\frac{n}{N}$ also serves as a preliminary estimate of task difficulty. Instances solved consistently by the baseline agent are regarded as relatively easy, while those with low success rates are more likely to require stronger search, evidence synthesis, and reasoning capabilities. We therefore prioritize difficult candidates for subsequent verification, particularly those satisfying $n$ $\leq$ $\frac{N}{2}$.

\subsubsection{Stage-III Verification: Trajectory Synthesis with a Strong Agent}
In the third stage, we employ a stronger LLM with agentic search tools to verify the difficult candidates retained from Stage II. For each selected QA pair, the strong agent independently attempts to solve the question $N$ times. We denote by $n$ the number of successful attempts for which the agent's final answer matches the reference answer.

If $n>0$, the QA pair is considered both challenging and verifiable. At least one successful trajectory indicates that the reference answer can be recovered from external evidence through a valid search and reasoning process, while also providing empirical support for the consistency of the question--answer mapping. We therefore retain the instance and preserve its successful trajectories as candidate supervision data. If $n=0$, the instance is discarded, as the pipeline cannot reliably establish that the question is solvable or that the reference answer is adequately supported.

This stage serves two complementary purposes. First, it filters out invalid, ambiguous, or unverifiable hard cases that cannot be resolved even by a stronger search agent. Second, it synthesizes high-quality, long-horizon search trajectories that can subsequently be used for supervised fine-tuning, reward-model training, and trajectory-level analysis.

\subsubsection{Trajectory Selection}
A single QA instance may admit multiple successful solution trajectories, but these trajectories are not necessarily equally useful for training. Some trajectories may contain redundant searches, unnecessary detours, or reasoning patterns that are highly similar to those of other trajectories. We therefore introduce a final selection stage in which a LLM serves as a trajectory judger.

For each QA pair, the LLM judger evaluates the successful trajectories according to four criteria: correctness, search efficiency, evidence grounding, and policy diversity. We retain a small subset of trajectories that are concise, well supported by external evidence, and meaningfully distinct from one another. Preference is given to trajectories that exhibit clear search intent, reliable evidence acquisition, and minimal irrelevant exploration, while preserving alternative strategies when multiple valid solution paths exist.

The final verified dataset contains QA instances that satisfy three conditions: they cannot be easily solved through parametric knowledge alone, they are answerable through external search, and they are supported by one or more high-quality agentic solution trajectories. This multi-stage verification and selection procedure improves the reliability, difficulty, and training utility of the synthesized task set.

\section{Agentic Post-training Pipeline}
Our post-training pipeline comprises two sequential stages: supervised fine-tuning (SFT) followed by reinforcement learning (RL). Both stages are constructed from the same underlying dataset, but adopt distinct data curation and sampling strategies. SFT teaches the model to invoke tools appropriately, interpret intermediate observations, and reason about the current state with respect to the overall task objective. RL subsequently refines the learned policy, improving both task performance and behavioral stability during long-horizon search trajectories.

\subsection{Tool Environment}
We equip the search agent with two tools: \textbf{search} and \textbf{visit}. Given a query, the \textbf{search} tool retrieves relevant online results, including titles, URLs, timestamps, and short snippets. The \textbf{visit} tool fetches the content of a target webpage and summarizes it under an explicit information-extraction objective, retaining evidence that is relevant to the agent's current goal.

To ground post-training in realistic interaction conditions, we construct a real-world tool environment backed by production-grade web services. Specifically, we use Serper\footnote{\url{https://www.serper.dev}} as the search backend, Jina\footnote{\url{https://jina.ai}} for webpage crawling, and Qwen3.5-27B for goal-conditioned content summarization. Although this configuration closely reflects real-world deployment, it introduces substantial challenges for post-training, where tool availability, response consistency, and execution stability are essential.

Production APIs are inherently volatile. They may exhibit variable latency, transient request failures, rate and concurrency limits, dynamically changing webpage content, and interference from advertisements or search-engine-optimized content. These uncertain sources can disrupt trajectory construction, reduce the reproducibility of training examples, and introduce considerable variance into online RL.

To mitigate these issues, we develop a robust tool-environment system built around three mechanisms: caching, disaster recovery, and quality enhancement. Together, they expose a stable and consistent tool interface to the agent throughout the post-training process.

\paragraph{Caching Mechanism}

The search tool employs query-level caching based on exact matching. Each live search stores the query and its corresponding results in Elasticsearch, and subsequent requests containing the identical query are served directly from the cache. By contrast, the fetch-and-summarize tool caches only the webpage crawling stage, using the URL as the cache key. For each requested URL, a cache hit returns the previously crawled webpage content, while a cache miss triggers a new crawl and stores the resulting content for future use. Goal-conditioned summaries are not cached. Consequently, the same webpage can produce different summaries under different information-seeking goals, while each URL is crawled at most once.

\paragraph{Disaster Recovery} 
The disaster-recovery mechanism maintains tool availability when external services experience high latency, rate limiting, or temporary outages, thereby preventing high-concurrency workloads from substantially reducing overall system throughput.

\paragraph{Quality Enhancement}
We apply several internal rules to improve the quality and robustness of tool responses. Specifically, advertisements are filtered from search results, and quotation marks are removed from search queries because they often cause the search API to return empty results.

\subsection{Supervised Fine-tuning}
We construct a ReAct-style scaffold using the available tool set and employ several closed- and open-source models as teachers for trajectory generation, including Qwen3.5-397B, Qwen3.6-Plus and Qwen3.7-max. The maximum context length is set to 256K tokens, without applying additional context-management strategies. To minimize inconsistencies in tokenization and trajectory formatting, all teacher models are selected from the same model family. The teachers are prompted to generate complete ReAct trajectories comprising intermediate reasoning, tool interactions, supporting evidence, and the final answer.

For each query in our SFT dataset, we perform $K$ independent rollouts and retain only trajectories whose final answers are judged to be correct by DeepSeek-V4-Pro.
The retained trajectories then undergo a series of quality-control filters, discarding trajectory that (i) contains fewer than 3 or more than 100 interaction turns, (ii) exhibits a failed tool-call rate above 20\%, (iii) shows a mismatch between the query language and the response language (e.g., an English response to a Chinese query), or (iv) contains repetitive patterns in its reasoning content. The resulting high-quality trajectories constitute the final corpus for supervised fine-tuning.



\subsection{Reinforcement Learning}
\subsubsection{RL Algorithm}
We adopt the Decoupled Clip and Dynamic sAmpling Policy Optimization (DAPO) algorithm \cite{yu2026dapo} to train our model. For each query-answer pair $(q, a)$ sampled from the data distribution $\mathcal{D}$, the policy is optimized by maximizing the following objective:
\begin{equation}
\begin{aligned}
\mathcal{J}_{\text{DAPO}}(\theta) ={}& \mathbb{E}_{(q,a) \sim \mathcal{D},\, \{o_i\}_{i=1}^{G} \sim \pi_{\theta_{\text{old}}}(\cdot \mid q)} \\
& \left[ \frac{1}{\sum_{i=1}^{G} |o_i|} \sum_{i=1}^{G} \sum_{t=1}^{|o_i|} \min\!\left( r_{i,t}(\theta)\, \hat{A}_{i,t},\; \text{clip}\!\left( r_{i,t}(\theta),\, 1 - \varepsilon_{\text{low}},\, 1 + \varepsilon_{\text{high}} \right) \hat{A}_{i,t} \right) \right] \\
\text{s.t.} \quad & 0 < \left| \left\{ o_i \mid \texttt{is\_equivalent}(a, o_i) \right\} \right| < G,
\end{aligned}
\end{equation}

where the importance sampling ratio and the group-relative advantage are defined as
\begin{equation}        
r_{i,t}(\theta) = \frac{\pi_\theta(o_{i,t} \mid q, o_{i,<t})}{\pi_{\theta_{\text{old}}}(o_{i,t} \mid q, o_{i,<t})}, \qquad
\hat{A}_{i,t} = \frac{R_i - \text{mean}(\{R_i\}_{i=1}^{G})}{\text{std}(\{R_i\}_{i=1}^{G})}.
\end{equation}

Here $\hat{A}_{i,t}$ is the normalized advantage shared across all tokens of response $o_i$, and $R_i$ denotes the reward assigned to the $i$-th response.
We use the decoupled clipping range where $\varepsilon_{\text{low}}=0.2$ and $\varepsilon_{\text{high}}=0.28$.

\subsubsection{Rollout Trajectory Analysis System}
In our experiments, we developed an automated trajectory analysis system to inspect reinforcement learning rollouts and guide the refinement of reward functions, training strategies, and experimental configurations. The system consists of two complementary components: (1) a self-evolving Judge Auditor that detects unreliable evaluation decisions and potential reward-hacking behaviors, and (2) a Tool-Behavior Diagnostic System that compares positive and negative trajectories with respect to tool-call validity, repetitive behavior, failure patterns, truncation, and context utilization.

\begin{figure}[!t]
    \centering
    \includegraphics[width=1.0\linewidth]{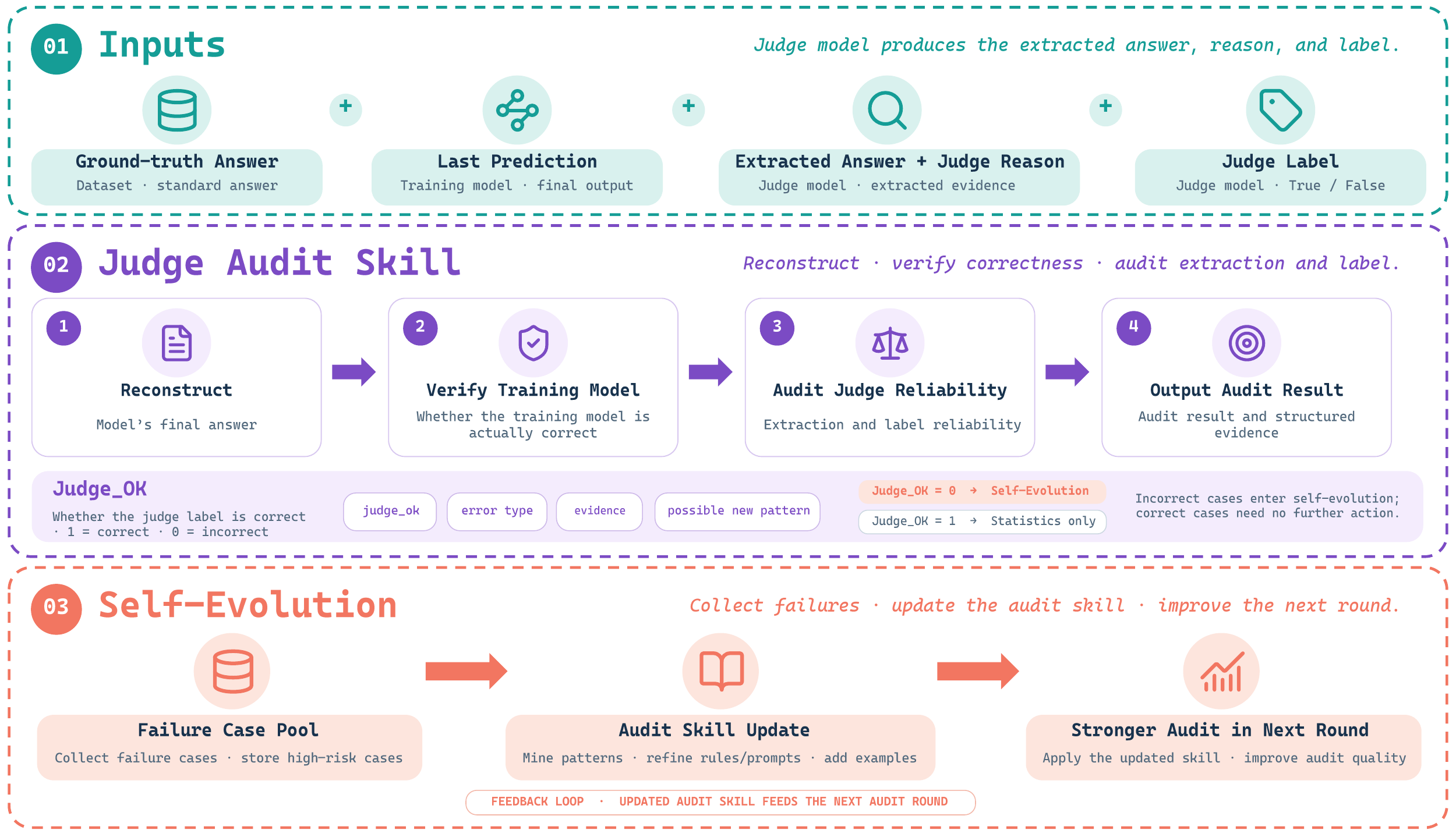}
    \caption{Overview of the Self-Evolving Judge Auditor.}
    \label{fig:judge_audit}
\end{figure}

\textbf{Self-Evolving Judge Auditor.}
This module identifies potential reward-hacking behaviors to improve training stability and the reliability of reward signals. Because policy behaviors evolve throughout training and may differ substantially across experimental configurations, it is impractical to enumerate all hacking patterns using fixed rules. We therefore introduce a self-evolving auditing mechanism that maintains an external \emph{Audit Skill}, which consolidates auditing criteria, known failure patterns, decision rules, and representative examples. After each experiment, newly identified errors and reward-hacking strategies are incorporated into the Audit Skill, thereby forming a closed-loop process of trajectory auditing, failure discovery, and skill refinement, as illustrated in Fig. \redref{fig:judge_audit}.
To ensure reliable evolution, we construct a stress-test suite consisting of three components: a frozen regression set containing previously resolved failures, an adversarial set comprising manually designed or LLM-generated hacking cases, and a clean control set of normal trajectories. An updated Audit Skill is accepted only if it achieves at least $95\%$ recall on hacking cases, maintains a false-positive rate below $5\%$ on clean samples, obtains a Macro-F1 score above $90\%$, and introduces no more than a one-percentage-point performance regression relative to the latest stable version. Otherwise, the update is rolled back, revised based on the identified failure cases, and subjected to another round of evaluation.

\begin{figure}[!t]
\centering
\includegraphics[width=1.0\linewidth]{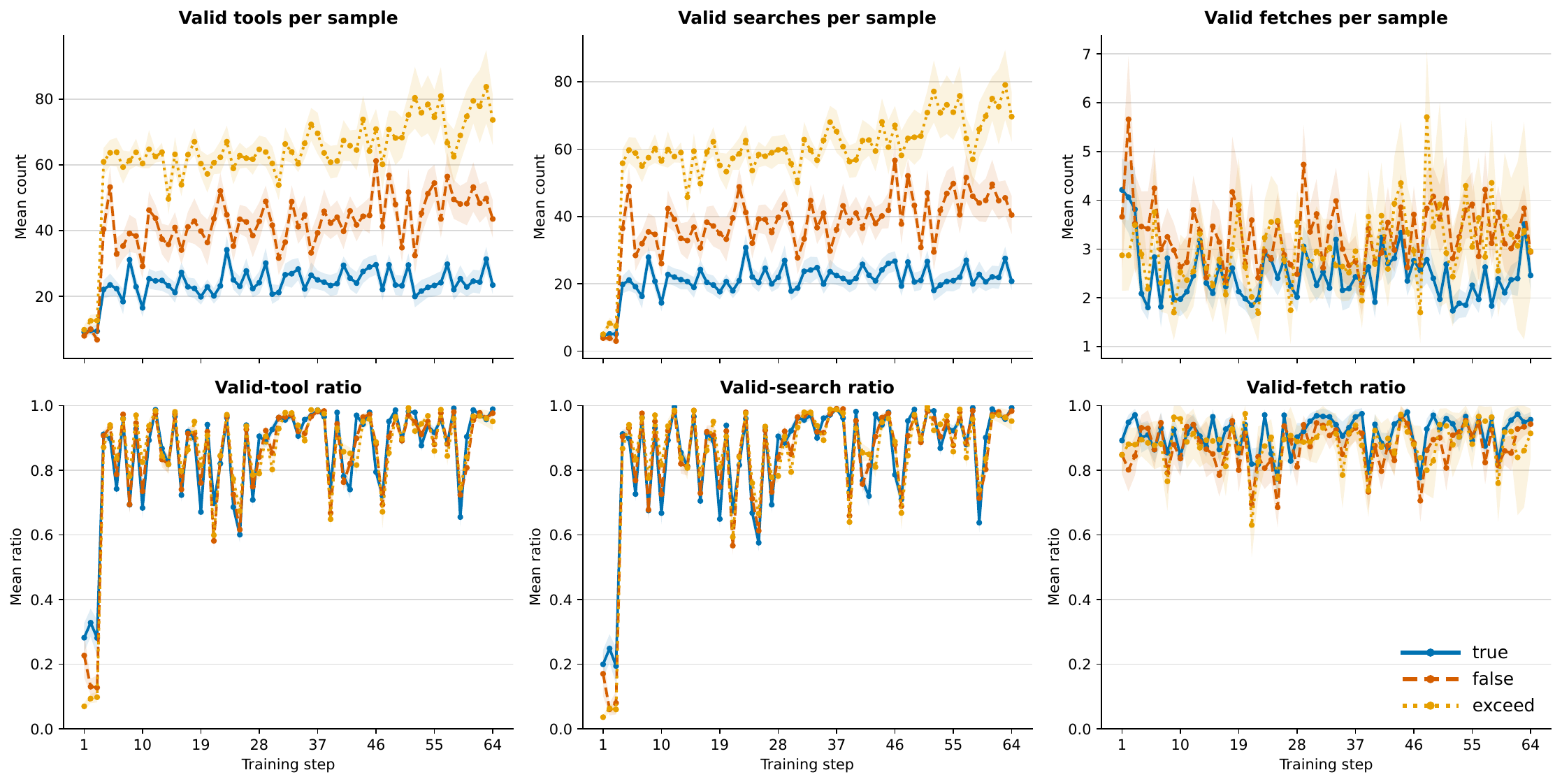}
\caption{Tool-use curves over training steps for rollout trajectories. A valid tool call is defined as a successfully executed invocation that returns a meaningful result. The top three plots show the average number of valid tool calls, while the bottom three show the proportion of valid tool calls.}
\label{fig:combine}
\end{figure}

\textbf{Tool-Behavior Diagnostic System}.
Beyond diagnosing reward-hacking behaviors at the outcome level, we develop a trajectory-level diagnostic system to systematically analyze tool-use behaviors. The system focuses on two major categories of problematic rollouts: overlength trajectories that exceed the context-window limit without producing a valid final answer, and trajectories that are ultimately judged to be unsuccessful. The system integrates LLM-based semantic analysis, tool-usage statistics, and rollout-tree structural analysis to enable fine-grained attribution of trajectory failures and to provide actionable guidance for optimizing training.
For example, as illustrated in Fig. \redref{fig:combine}, comparisons of the number and proportion of valid tool calls across trajectory categories reveal a persistent \emph{over-searching} pattern in the policy model. In particular, many overlength trajectories arise from excessive, repetitive, and low-yield search behavior, rather than from failures in the external environment. Motivated by this finding, we introduce a behavior-specific penalty for redundant and unproductive tool use, rather than simply masking all overlength samples during training. This strategy retains useful supervision from long yet informative trajectories while explicitly discouraging inefficient search loops.

\subsection{Composite Reward Design}
In long-horizon agentic systems, performance depends on both the correctness of the final answer and the quality of the intermediate reasoning and tool-use trajectory. Therefore, we employ a hybrid reward modeling framework that combines \textbf{outcome-level supervision} and \textbf{process-level supervision}. The overall reward is formulated as:
\begin{equation}
R = R_{\text{outcome}} + \lambda_1 R_{\text{format}} + \lambda_2 R_{\text{turn}}
\end{equation}
where $R_{\text{outcome}}$ measures semantic correctness of the final answer, while process rewards $R_{\text{format}}$ and $R_{\text{turn}}$ regularize structured tool usage and interaction efficiency.

Compared to purely outcome-driven approaches, this decomposition offers several key advantages:
\begin{itemize}[leftmargin=1em, listparindent=\parindent]
\item \textbf{Improved Credit Assignment}: Process-level rewards provide dense, step-wise feedback throughout the trajectory, which reduces the variance of policy gradient updates and leads to more stable optimization.

\item \textbf{Tool-use Reliability}: Format constraints enforce strict compliance with tool invocation schemas, preventing malformed operations that would otherwise cause execution failures.

\item \textbf{Compute Efficiency Control}: Turn-level rewards explicitly regulate the reasoning budget, discouraging excessively long interaction loops and mitigating unbounded or redundant information-gathering behavior.
\end{itemize}

Overall, this hybrid reward architecture enables stable reinforcement learning in complex multi-step agent environments where both correctness and interaction quality are critical.

\subsubsection{Outcome Reward}
The outcome reward follows \textbf{rule-first, judge-fallback} paradigm that combines deterministic rule-based verification with LLM-based semantic evaluation. Rule-based evaluation provides high-precision, deterministic correctness but is limited by the expressiveness of hand-crafted rules. In contrast, LLM-based evaluation provides flexible semantic evaluation, enabling robust assessment of paraphrases, multilingual expressions, and other semantically equivalent variations, albeit with lower reward stability.


Given a predicted answer $y$ and reference solution $y^{*}$, we first apply deterministic rule-based verification. These rules include exact match, numerical consistency, set equality, or other task-specific symbolic validations. Formally:
\begin{equation}
R_{\text{rule}} =
\begin{cases}
1, & \text{if } \text{Verify}_{\text{rule}}(y, y^{*}) = \text{True} \\
\varnothing, & \text{if rule-based verification is inconclusive}
\end{cases}
\end{equation}
The verifier returns an \emph{inconclusive} result when deterministic rules cannot reliably determine correctness. This typically occurs due to (i) semantically equivalent paraphrases, (ii) entity aliasing (e.g., multiple valid names for the same entity), or (iii) multilingual and formatting variations.
In such cases, an LLM judge is invoked as a fallback to evaluate the correctness of the predicted answer and the reference:


\begin{equation}
R_{\text{judge}} = \text{LLM}(query, y, y^{*}) \in [0, 1]
\end{equation}





\subsubsection{Process Reward}

While outcome rewards supervise final answer correctness, they provide limited guidance on intermediate reasoning and tool usage quality. To address this, we introduce process-level rewards that explicitly regularize both \textbf{tool invocation validity} and \textbf{interaction efficiency} in agent trajectories.

\paragraph{Format Reward for Structured Tool Calls}
Agent trajectories rely on structured tool calls (e.g., XML/JSON). Even minor format violations lead to execution failure and invalidate the entire rollout. Therefore, we model format correctness as a \textbf{hard constraint}.
Let $c_t$ denote the tool call at step $t$, with validity indicator $\text{Valid}(c_t) \in \{0,1\}$.
The format reward is defined at the trajectory level as:

\begin{equation}
R_{\text{format}} =
\begin{cases}
1, & \text{if } \forall t \in [1, T], \text{Valid}(c_t) = 1 \\
0, & \text{otherwise}
\end{cases}
\end{equation}
The validity function evaluates whether each tool invocation conforms to the expected protocol, including XML/JSON syntax correctness, the presence of required fields, the consistency of argument types, and compliance with the tool's schema specification.

\paragraph{Turn-level Efficiency Reward}
We introduce a turn-level reward that models the non-monotonic relationship between trajectory length and reasoning quality. Empirically, agent performance does not improve monotonically with longer trajectories: insufficient reasoning causes premature termination, while excessive interaction leads to inefficiency and search looping. To capture this, we define a non-monotonic utility function over the normalized computation budget. 

Let $N$ denote the number of interaction turns in a rollout and $N_{\max}$ be the maximum allowed. We define the normalized trajectory length as:
\begin{equation}
x = \frac{N}{N_{\max}} \in [0,1]
\end{equation}
We introduce a piecewise smooth reward parameterized by four thresholds $\tau_1 < \tau_2 < \tau_3 < \tau_4 \in (0, 1)$,  which partition the budget into five behavioral regimes:

\begin{equation}
R_{\text{turn}} =
\begin{cases}
-1 + \left(\frac{x}{\tau_1}\right)^2, & x < \tau_1 \\
\frac{x - \tau_1}{\tau_2 - \tau_1}, & \tau_1 \le x < \tau_2 \\
1, & \tau_2 \le x \le \tau_3 \\
1 - \frac{x - \tau_3}{\tau_4 - \tau_3}, & \tau_3 < x \le \tau_4 \\
- \left(\frac{x - \tau_4}{1 - \tau_4}\right)^2, & x > \tau_4
\end{cases}
\end{equation}
The five regimes carry distinct semantics:
\begin{itemize}[leftmargin=1em, listparindent=\parindent]
\item \textbf{Under-thinking ($x < \tau_1$)}: Trajectories terminate too early with insufficient reasoning depth. The quadratic penalty strongly discourages extremely short interactions while providing a smooth gradient toward increased exploration.

\item \textbf{Warm-up ($\tau_1 \le x < \tau_2$)}: A linear ramp incentivizes the agent to reach a minimally viable reasoning budget, serving as a transition from under-thinking to stable behavior.

\item \textbf{Optimal Plateau ($\tau_2 \le x \le \tau_3$)}: The reward is maximized and length-invariant, reflecting that trajectories within this range are computationally efficient and sufficiently thorough.

\item \textbf{Decay ($\tau_3 < x \le \tau_4$)}: Additional steps yield diminishing returns. The linear decrease penalizes over-computation without abrupt discontinuity.

\item \textbf{Over-thinking / looping ($x > \tau_4$)}: A quadratic penalty strongly discourages pathological looping behaviors such as repetitive reasoning cycles or redundant searches, which consume budget without meaningful progress.
\end{itemize}

Unlike monotonic length penalties or symmetric U-shaped rewards, this design introduces a \textbf{stable optimal plateau} bounded by smooth transition regions. This structure offers two key benefits for optimization: (i) the flat plateau improves policy stability by reducing sensitivity to small fluctuations in trajectory length, and (ii) smooth boundary transitions ensure gradient continuity, enabling stable reinforcement learning dynamics in long-horizon agentic settings.

\section{SearchArt Harness Design}

SearchArt implements a harness design that formulates agentic search as an iterative reasoning-and-search process. At each iteration, the agent interleaves internal reasoning with external web-tool invocations to gather evidence, manage working context, and refine its response. To improve robustness and reliability, the harness integrates context management and self-evaluation mechanisms. We instantiate this design in two variants tailored to deep search and deep research scenarios, respectively.

\subsection{Tool Settings}
We equip the search agent with three external tools—\textsc{Search}, \textsc{Visit}, and \textsc{Check}—which together define the interaction interface between the agent and the real-world web environment.

\paragraph{Search.} The \textsc{Search} tool accepts a natural-language query as input and returns a ranked list of online search results. Each result contains structured metadata including the publication or retrieval time, page title, URL, and a brief snippet. 
\paragraph{Visit.} The \textsc{Visit} tool takes a target URL and a task-specific goal as input, and returns a condensed summary of the webpage content conditioned on the given goal. 
\paragraph{Check.} The \textsc{Check} tool evaluates whether the retrieved evidence sufficiently supports a candidate answer by assigning a verification confidence score from 0 to 100. We introduce this tool because LLMs may resort to unsupported speculation after repeated unsuccessful reasoning and retrieval attempts. The agent is allowed to finalize an answer only when the verification score exceeds 90. Otherwise, it must continue searching for additional evidence or abstain from providing a definitive answer. Thus, \textsc{Check} serves as an explicit evidence-sufficiency gate, improving the reliability and groundedness of the final response.

To approximate realistic deployment conditions, we use a production-level web environment. Specifically, \textsc{Search} is implemented using Serper\footnote{\url{https://www.serper.dev}}, and \textsc{Visit} is implemented using Jina\footnote{\url{https://jina.ai}} for webpage crawling, with Qwen3.5-27B serving as the summarization model for extracting key webpage information. This setup ensures that the agent interacts with real online information sources rather than a closed or pre-indexed benchmark environment.

\subsection{DeepSearch}
DeepSearch adopts a ReAct-based scaffold that interleaves iterative reasoning with external tool invocations \cite{yao2022react}. To mitigate context accumulation and error propagation, it employs a discard-all context management strategy.

\subsubsection{ReAct Trajectory}
The ReAct (Reasoning-Acting-Observation) trajectory consists of a sequence of steps, each containing a reasoning state, a tool action, and the corresponding observation. The agent updates its state after each interaction until producing a final answer. These trajectories capture the full search process and support supervised fine-tuning, verification, and process-level evaluation.

\subsubsection{Context Management}
Even with a long context window, search-based agent trajectories can easily exceed the model's context limit due to repeated tool calls, lengthy webpage contents, accumulated notes, and redundant evidence. Following DeepSeek \cite{xu2026deepseek}, we adopt the \texttt{Discard-all} strategy. When the context length exceeds a predefined fraction of the maximum budget, we reset the context by discarding all the previous tool call history and restart the agent rollout. This mechanism is especially effective for challenging DeepSearch tasks that require extensive search and evidence gathering.

\subsection{DeepResearch}
\label{subsec:deepresearch}

We introduce \textbf{DeepResearch}, a harness designed for automated research tasks, built upon a dual-space architecture and a decoupled planning-writing paradigm.


\subsubsection{Motivation \& Design}

Existing research agents based on the ReAct paradigm typically accumulate tool calls and observations within a single context window before generating the final report.  While effective for short question-answering tasks, it becomes unreliable for long-horizon, information-intensive research tasks due to two major limitations:
\begin{itemize}[leftmargin=1em, listparindent=\parindent]
    \item \textbf{Token Overload Risk.} As tool calls and observations continuously accumulate, the context grows substantially and may approach the model's maximum context capacity, eventually interrupting the research process.
    \item \textbf{Decision Quality Degradation.} The increasing amount of irrelevant or redundant information dilutes useful evidence available for decision-making and degrade the quality of subsequent reasoning steps. This may further lead to  logical inconsistencies, citation misalignment, and unreliable conclusions in the final report.
\end{itemize}
To address these limitations, DeepResearch follows the principle of \textit{exploration first, creation second, with clear separation of concerns}. It decomposes the research process into two stages: a \textbf{planning phase} for broad information gathering and knowledge system construction, and a \textbf{writing phase} for focused in-depth analysis and report organization. Through this decoupling design, DeepResearch reduces reasoning entropy and improves the quality, reliability, and efficiency of long-form research report generation.

\subsubsection{Overall Architecture}

A key architectural feature of DeepResearch is its \textbf{dual-space} design, which explicitly separates active decision-making from persistent knowledge storage. This separation allows the primary agent to operate within a compact, low-noise reasoning context while retaining detailed research artifacts in an auxiliary knowledge space. As a result, DeepResearch preserves evidence coverage and citation traceability without overloading the active context, thereby improving the stability and reliability of long-horizon research generation.

\paragraph{Hierarchical Memory System.}

This dual-space architecture is implemented through a hierarchical memory system consisting of two complementary layers that interact dynamically throughout the research workflow.
\textbf{Primary Working Memory} maintains condensed information, including webpage summaries, structured outline nodes, key findings, etc. It provides a low-noise reasoning context to guarantee agent core decisions such as planning, query formulation, research direction refinement.
\textbf{Auxiliary Knowledge Memory} stores full research artifacts, including raw webpage contents, citation mappings, and evidence indexes. It ensures information persistence and citation traceability while remaining transparent to the agent's active reasoning process.

\subsubsection{Planner Phase: Outline Construction and Evidence Accumulation}
The planner phase is the \textit{outward exploration} component. Its core goal is to perform global information collection, construct a structured research outline and deposit a tracable evidence library for the subsequent writing phase. It adopts an \textit{exploration-review-iteration} cycle with the following evolving states: 

\paragraph{Evidence Retrieval.} The agent acquires new evidence through search and browsing actions, progressively expanding its understanding of the research topic.

\paragraph{Synchronous Memory Update.} 
The collected evidence undergoes URL-based deduplication and quality filtering to remove invalid sources, such as login-restricted pages, permission-denied content, and other unusable webpages. The retained evidence is integrated into the auxiliary memory, while only concise, decision-relevant summaries are selectively injected into the primary memory context, to maintain contextual efficiency and reduce unnecessary information overhead.

\paragraph{Outline Evolution.}
The agent refines the outline based on newly accumulated evidence, updating both the structure and the citations. Following each update, an automated integrity review is performed to identify potential information gaps. If insufficient coverage or evidence support is detected, the agent initiates another retrieval cycle to further acquire evidence and improve the outline.

The planner terminates when the outline reaches sufficient coverage and evidence support, subject to predefined iteration constraints, including both minimum and maximum limits. These controls balance exploration depth with computational efficiency, ensuring sufficient information discovery while avoiding unnecessary retrieval overhead.
We give an example of hierarchical research outline in Appendix \redref{appendix:DeepResearch Outline}

\subsubsection{Writer Phase: Parallel Chapter-Wise Writing and Full-Text Integration}
The writer phase is the \textit{inward digging} component of the framework. Its core goal is to complete high-quality report writing and standardized reference output based on the accumulated structured evidence. In this phase, the context is completely reset to fully isolate information noise from the planner phase.

\paragraph{Parallel Chapter-wise Writing.}
To improve long-form generation efficiency, the framework adopts a context-isolated parallel writing strategy. 
Each chapter is generated independently with a dedicated instruction defining its writing scope, including the original research request, its chapter outline, and the corresponding filtered evidence identified by outline citation tags. This design minimizes irrelevant context interference and significantly reduces the overall generation time for long reports.

\paragraph{Full-Text Integration and Reference Generation.}
After completing chapter-wise drafting, the system performs full-text integration and standardized output. The agent takes all chapter drafts as input and performs logical cohesion, style unification and full-text polishing. 
The reference section is generated through codes from evidence metadata, ensures standardized formatting and a one-to-one mapping between citation numbers and evidence.

\paragraph{Long-Text Generation Guarantee.}
The framework provides an automatic output continuation mechanism that detects incomplete generations caused by length limits and seamlessly resumes generation. This enables the production of research reports of arbitrary length without manual intervention.

\section{Experiments}
\subsection{Experimental Setup}

\paragraph{Training Setup.}
We adopt Qwen3.5-27B as the base model. 
For supervised fine-tuning, we optimize the model with standard cross-entropy loss while masking all the tool-response tokens from loss computation. We train the model with the AdamW optimizer, using an inital learning rate of $2 \times 10^{-6}$ that follows a cosine decay scheduler with a mininum learning rate of $2 \times 10^{-7}$. The maximum training sequence length is set to 128,000 tokens.
For reinforcement learning, we use VERL framework and adopt the DAPO algorithm. Empirically, we set the turn-penalty hyperparameters to $\tau_1=0.1$, $\tau_2=0.2$, $\tau_3=0.7$, and $\tau_4=0.8$. The actor learning rate is set to $1 \times 10^{-6}$ with a mini-batch size of 32 and 8 sampled trajectories per prompt. The maximum context length is set to 128,000 tokens. During rollout, we use a temperature of 1.0 and allow up to 100 tool-call turns.
All experiments are performed on Atlas 900 A3 SuperPoD.

\paragraph{Evaluation Setup.}
We evaluate our model on five challenging search-oriented benchmarks: BrowseComp \cite{wei2025browsecomp}, BrowseComp-ZH \cite{zhou2025browsecomp}, BrowseComp-Plus \cite{chen2025browsecomp}, WideSearch \cite{wong2025widesearch}, and DeepResearch Bench \cite{du2025deepresearch}. 
This evaluation suite is designed to measure the model’s ability to retrieve relevant information from the web, integrate evidence from multiple sources, reason over long search trajectories, and produce accurate answers for complex information-seeking tasks in both English and Chinese.

For BrowseComp series benchmarks, we follow DeepSeek \cite{guo2025deepseek} and adopt a discard-all strategy.
For WideSearch, we use the official inference scaffold and evaluation script \cite{wong2025widesearch}.
For DeepResearch Bench, we adopt the method described in Section~\redref{subsec:deepresearch}, using a dual-space scaffold that separates the agent’s compact decision context from a persistent full-evidence store while iteratively constructing a citation-grounded outline through search and evidence accumulation. The final report is generated by independently writing outline-aligned chapters with filtered evidence, followed by full-text integration and programmatic reference generation.

\begin{table}[!t]
\small
\centering
\caption{Performance comparison across various agentic benchmarks. $^*$: Reproduced by our implementation. $^\dagger$: Reported in the official benchmark report (if a context management strategy is used). $^\#$: Reproduced by third-party unofficial reports.}
\label{tbl:mian_results} 
\begin{tabular}{l c c c c c c}
\toprule
\textbf{Model} & \textbf{Size} & \textbf{\makecell{Browse\\-Comp}} & \textbf{\makecell{BrowseComp\\-ZH}} & \textbf{\makecell{BrowseComp\\-Plus}} & \textbf{\makecell{Wide\\-Search}} & \textbf{\makecell{DeepResearch\\Bench}} \\ 
\midrule
\multicolumn{7}{l}{\textbf{Closed-Source}} \\
\midrule
GPT 5.5 & - & 84.4$^\dagger$ & - & - & - & -   \\
GPT 5.4 & - & 82.7$^\dagger$ & 75.3$^\dagger$ & - & - & -   \\
GPT 5.3 & - & 77.3$^\dagger$ & - & - & - & -   \\
GPT 5.2 & - & 65.8$^\#$ & 76.1$^\#$ & - & 76.8$^\#$ & 49.4$^\#$ \\
GPT 5 & - & 54.9$^\dagger$ & 63$^\dagger$ & 70.12$^\dagger$ & 62.2$^\dagger$ & - \\
Gemini 3.1 Pro & - & 85.9$^\dagger$ & 74.8$^\#$ & - & - & -  \\
Gemini 3.0 Pro & -  & 59.2$^\#$ & 66.8$^\#$ & - & 68.0$^\#$ &  - \\
Gemini 3.0 Flash & -  & 41.5$^\dagger$ & 63$^\dagger$ &  &  &    \\
Claude Opus 4.8 & - & 84.3$^\dagger$ & - & 74.34$^\dagger$ & - & -  \\
Claude Opus 4.7 & - & 79.3$^\dagger$ & - & - & - & -   \\
Claude Opus 4.6 & - & 84.0$^\dagger$ & - & - & - & -  \\
Claude Opus 4.5 & - & 67.8$^\#$ & 62.4$^\#$ & 70.48$^\dagger$/ 85.30$^\dagger$ & 76.4$^\#$ & -  \\
Seed 2.0 Pro-0215 & - & 77.3$^\dagger$ & 82.4$^\dagger$ & - & 74.7$^\dagger$ & -  \\
Seed 1.8 & - & 67.6$^\dagger$ & 81.3$^\dagger$ & - & 63.8$^\dagger$ & 53.3$^\#$  \\
\midrule
\multicolumn{7}{l}{\textbf{Open-Source ($\textgreater$ 700B)}} \\
\midrule
DeepSeek-V4-Pro & 1.6T-A49B & 83.4$^\dagger$ & - & - & - & -  \\
DeepSeek-V3.2 & 671B-A37B & 51.4$^\dagger$/ 67.6$^\dagger$ & 65.0$^\dagger$ & - & 32.5$^\dagger$ & 45.6$^\#$ \\
GLM 5.1 & 744B-A40B & 68.0$^\dagger$/ 79.3$^\dagger$ & - & - & -  & -   \\
GLM 5 & 744B-A40B &  62.0$^\dagger$/ 75.9$^\dagger$ & 72.7$^\dagger$ & - & - & 46.9$^\#$ \\
Kimi K2.6 & 1T–A32B & 83.2$^\dagger$ & - & - & 80.8$^\dagger$ & -  \\
Kimi K2.5 & 1T–A32B & 60.6$^\dagger$/ 74.9$^\dagger$ & 62.3$^\dagger$ & - & 72.7$^\dagger$ & 45.5$^\#$  \\
Kimi K2 & 1T–A32B & 60.2$^\dagger$ & 62.3$^\dagger$ & - & - & 45.65$^\#$  \\
\midrule
\multicolumn{7}{l}{\textbf{Open-Source (200B $\sim$ 400B)}} \\
\midrule
Qwen3.5-397B-A17B & 397B-A17B & 78.6$^\dagger$ & 70.3$^\dagger$ & 58.31$^*$ & 74.0$^\dagger$ & 51.26$^*$  \\
GLM 4.7 & 355B-A32B & 52.0$^\dagger$/ 67.5$^\dagger$  & 66.6$^\dagger$ & 50.18$^\dagger$ & 52.6$^*$ & 50.43$^*$  \\
DeepSeek-V4-Flash & 284B-A13B & 73.2$^\dagger$ & 60.21$^*$ & 62.44$^*$ & 51.8$^*$ & 54.28$^*$  \\
Minimax 2.5 & 230B-A10B & 76.3$^\dagger$ & 44.29$^*$ & 49.16$^*$ & 70.3$^\dagger$/ 53.0$^*$ & 51.05$^*$  \\
\midrule
\multicolumn{7}{l}{\textbf{Post-Training Search Agent}} \\
\midrule
MiroThinker-1.7-30B & 30B-A3B & 67.9$^\dagger$ & 72.3$^\dagger$ & - & - & - \\
MiroThinker-1.5-30B & 30B-A3B & 56.1$^\dagger$ & 66.8$^\dagger$ & - & 37.9$^*$ & - \\
OpenSeeker & 30B-A3B & 29.5$^\dagger$ & 48.4$^\dagger$ & - & 59.4$^\dagger$ & - \\
Tongyi-DeepResearch & 30B-A3B & 43.4$^\dagger$ & 46.7$^\dagger$ & - & 41.7$^*$ & - \\
RedSearcher & 30B-A3B & 57.4$^\dagger$ & 58.2$^\dagger$ & - & - & - \\
LongSeeker & 30B-A3B & 61.5$^\dagger$ & 62.5$^\dagger$ & - & - & - \\
S1-DeepResearch & 32B & 36.7$^\dagger$ & 48.4$^\dagger$ & - & - & 46.5$^\dagger$ \\
O-Researcher-72B & 72B & - & - & - & - & 48.48$^\dagger$ \\
\midrule
\multicolumn{7}{l}{\textbf{Our Base \& Post-Train}} \\
\midrule
Qwen3.5-27B & 27B & 61.0$^\dagger$ & 62.1$^\dagger$ & 58.43$^*$ & 61.1$^\dagger$ & 44.7$^*$  \\
\rowcolor{lightblue}
SearchArt-27B & 27B & 70.06$^*$ & 74.39$^*$ & 63.49$^*$ & 64.0$^*$ & 52.55$^*$ \\ 
\bottomrule
\end{tabular}
\label{table:main}
\end{table}

\subsection{Experimental Results}
\subsubsection{Overall Performance}
As shown in Table \redref{tbl:mian_results}, SearchArt-27B consistently outperforms its base model, Qwen3.5-27B, across all five benchmarks, demonstrating that the proposed post-training strategy substantially improves both deep-search and deep-research capabilities. On BrowseComp, SearchArt-27B improves the score from 61.00 to 70.06, corresponding to an absolute gain of 9.06 points and a relative improvement of 14.85\%. The score on BrowseComp-ZH increases from 62.10 to 74.39 (+12.29 points; +19.79\%), while BrowseComp-Plus improves from 58.43 to 63.49 (+5.06 points; +8.66\%). SearchArt-27B also raises the Wide-Search score from 61.10 to 64.0 (+2.9 points; +4.92\%) and the DeepResearch Bench score from 44.70 to 52.55 (+7.85 points; +17.56\%). Averaged over the five benchmarks, post-training produces an absolute improvement of 7.43 points, equivalent to a mean relative gain of 13.16\%.

Despite having only 27B parameters, SearchArt-27B achieves performance comparable to, and in several cases better than, substantially larger open-source models. Among models in the 200B–400B range, SearchArt-27B obtains 74.39 on BrowseComp-ZH, outperforming Qwen3.5-397B-A17B (70.30), GLM 4.7 (66.60), DeepSeek-V4-Flash (60.21), and MiniMax 2.5 (44.29). Its BrowseComp-Plus score of 63.49 also exceeds those of Qwen3.5-397B-A17B (58.31), GLM 4.7 (50.18), DeepSeek-V4-Flash (62.44), and MiniMax 2.5 (49.16). On DeepResearch Bench, SearchArt-27B reaches 52.55, surpassing Qwen3.5-397B-A17B (51.26), GLM 4.7 (50.43), and MiniMax 2.5 (51.05), while remaining close to DeepSeek-V4-Flash (54.28). It additionally achieves 64.0 on Wide-Search, exceeding GLM 4.7 (52.60) and DeepSeek-V4-Flash (51.80). These results indicate that SearchArt-27B can match the search and research capabilities of models that are approximately one order of magnitude larger.

SearchArt-27B also demonstrates competitive performance against open-source models with more than 700B parameters and several proprietary systems. On BrowseComp-ZH, its score of 74.39 exceeds those of GLM 5 (72.70), DeepSeek-V3.2 (65.00), Kimi K2.5 (62.30), and Kimi K2 (62.30). On DeepResearch Bench, its score of 52.55 is higher than DeepSeek-V3.2 (45.60), GLM 5 (46.90), Kimi K2.5 (45.50), and Kimi K2 (45.65). Compared with proprietary models, SearchArt-27B achieves 70.06 on BrowseComp, outperforming GPT-5.2 (65.80), GPT-5 (54.90), Gemini 3.0 Pro (59.20), Claude Opus 4.5 (67.80), and Seed 1.8 (67.60). Its BrowseComp-ZH result is also close to Gemini 3.1 Pro (74.80) and GPT-5.4 (75.30), with gaps of only 0.41 and 0.91 points, respectively. On DeepResearch Bench, SearchArt-27B surpasses GPT-5.2 (49.40) and approaches Seed 1.8 (53.30), whereas its Wide-Search score of 64.0 slightly exceeds GPT-5 (62.20) and Seed 1.8 (63.80).

Notably, the largest gains are observed on BrowseComp-ZH and DeepResearch Bench. BrowseComp-ZH exhibits the greatest improvement, increasing by 12.29 points or 19.79\%, suggesting that post-training is particularly effective at strengthening Chinese-language deep-search capabilities. DeepResearch Bench records the second-largest relative improvement, with a gain of 7.85 points or 17.56\%, indicating that the benefits extend beyond answer-oriented web search to more complex research workflows involving iterative evidence acquisition, information synthesis, and report-level reasoning. Together, these results demonstrate that SearchArt-27B achieves strong parameter efficiency and that the proposed post-training approach is especially beneficial for Chinese deep search and long-horizon deep research.

\subsubsection{Before \& After Post-Training Comparison}
\begin{figure}[!t]
    \centering
    \includegraphics[width=1\linewidth]{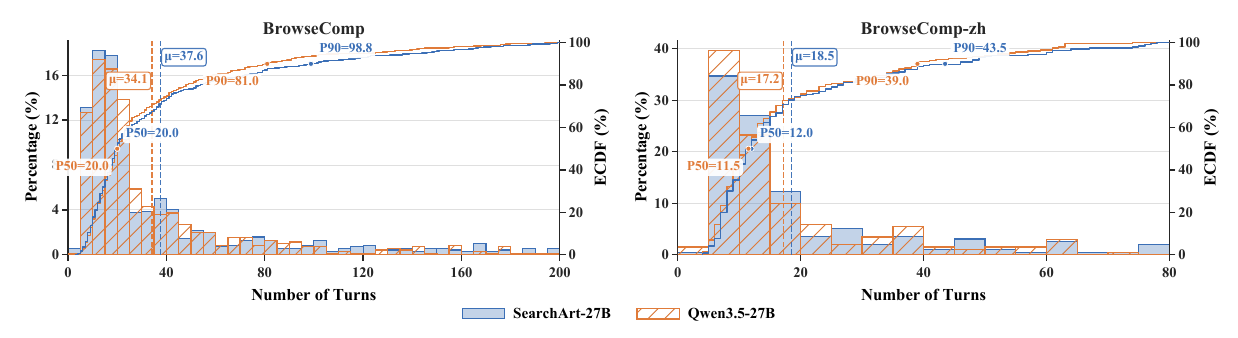}
    \caption{Distributions of the number of turns on BrowseComp (left) and BrowseComp-zh (right). Bars show the percentage of trajectories, while solid curves represent the ECDFs. Annotations indicate the mean (\(\mu\)), 50th percentile (P50), and 90th percentile (P90).}
    \label{fig:post-train-behavior}
\end{figure}
\begin{figure}[!t]
    \centering
\includegraphics[width=1\linewidth]{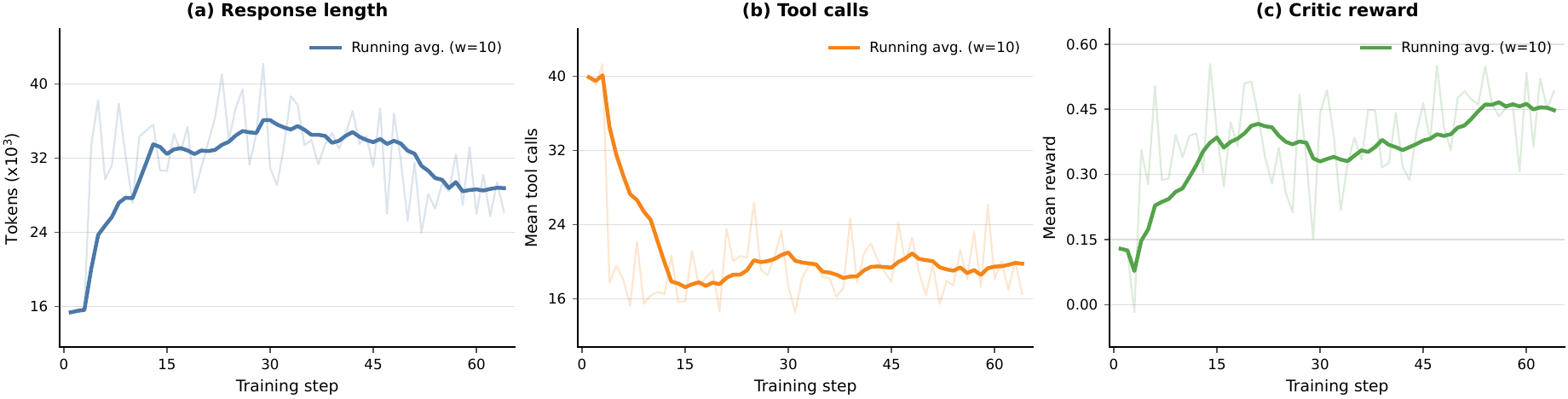}
    \caption{Training dynamics of SearchArt-27B: (a) mean response length, (b) mean number of tool calls, and (c) mean critic reward.}
    \label{fig:post-train-process}
\end{figure}

Figure \redref{fig:post-train-behavior} highlights the more persistent search behavior of SearchArt-27B model. On both datasets, the ECDF of SearchArt-27B is generally shifted to the right of that of Qwen3.5-27B, indicating that SearchArt-27B is more willing to continue searching instead of terminating a trajectory prematurely. This difference is particularly pronounced on BrowseComp, where the mean turn count increases from 34.1 for Qwen3.5-27B to 37.6 for SearchArt-27B. Although both models have a median of 20 turns, the 90th percentile increases from 81.0 to 98.8 turns, showing that the additional search effort is concentrated primarily on difficult, long-running trajectories rather than being uniformly applied to all examples. A similar pattern is observed on BrowseComp-zh: SearchArt-27B has a mean of 18.5 turns compared with 17.2 for Qwen3.5-27B, while their 90th percentiles are 43.5 and 39.0 turns, respectively. These results suggest that fine-tuning enables SearchArt-27B to preserve a similar computational profile on typical cases while allocating a larger search budget to challenging instances. Such selective persistence is beneficial for agentic search tasks, where additional exploration can help gather more complete evidence and reduce the risk of premature termination.

\subsubsection{Training Process}
We show several key training dynamics in Figure \redref{fig:post-train-process}. The critic reward steadily improves throughout training, indicating that the policy learns to produce more successful trajectories. Meanwhile, the mean number of tool calls drops sharply during the early stage and subsequently stabilizes, suggesting that the agent learns a more efficient tool-use strategy. Response length initially increases as the model explores longer reasoning and tool-interaction trajectories, then gradually decreases and stabilizes, consistent with a transition toward more concise yet effective solutions.

\begin{figure}[!t]
    \centering
    \begin{subfigure}[t]{0.7\linewidth}
        \centering
        \includegraphics[width=\linewidth]{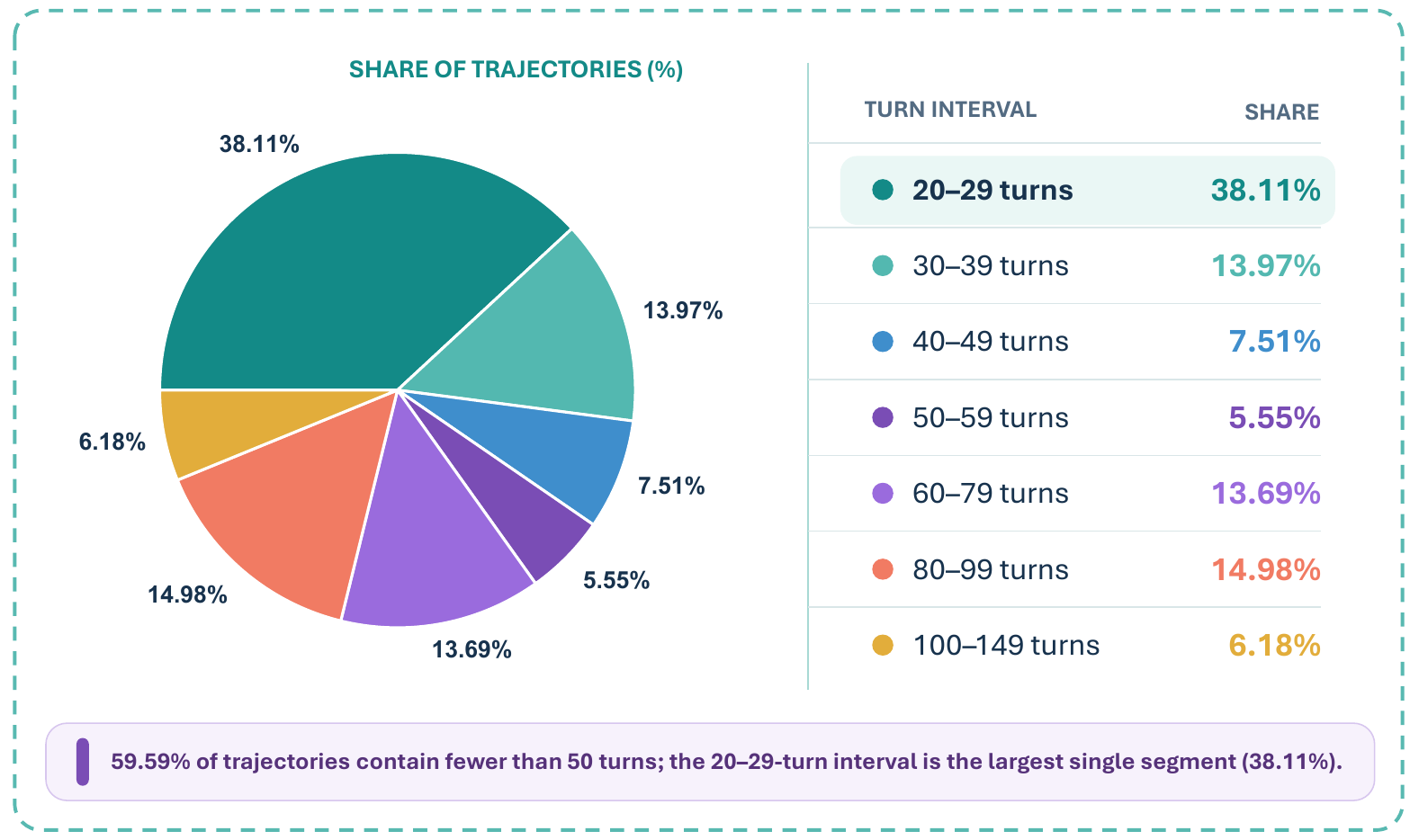}
        \caption{Trajectory turn distribution of DeepSearch QA.}
        \label{fig:synthetic_task_qa_trajectory_distribution_bright}
    \end{subfigure}
    \hfill
    \begin{subfigure}[t]{0.7\linewidth}
        \centering
        \includegraphics[width=\linewidth]
        {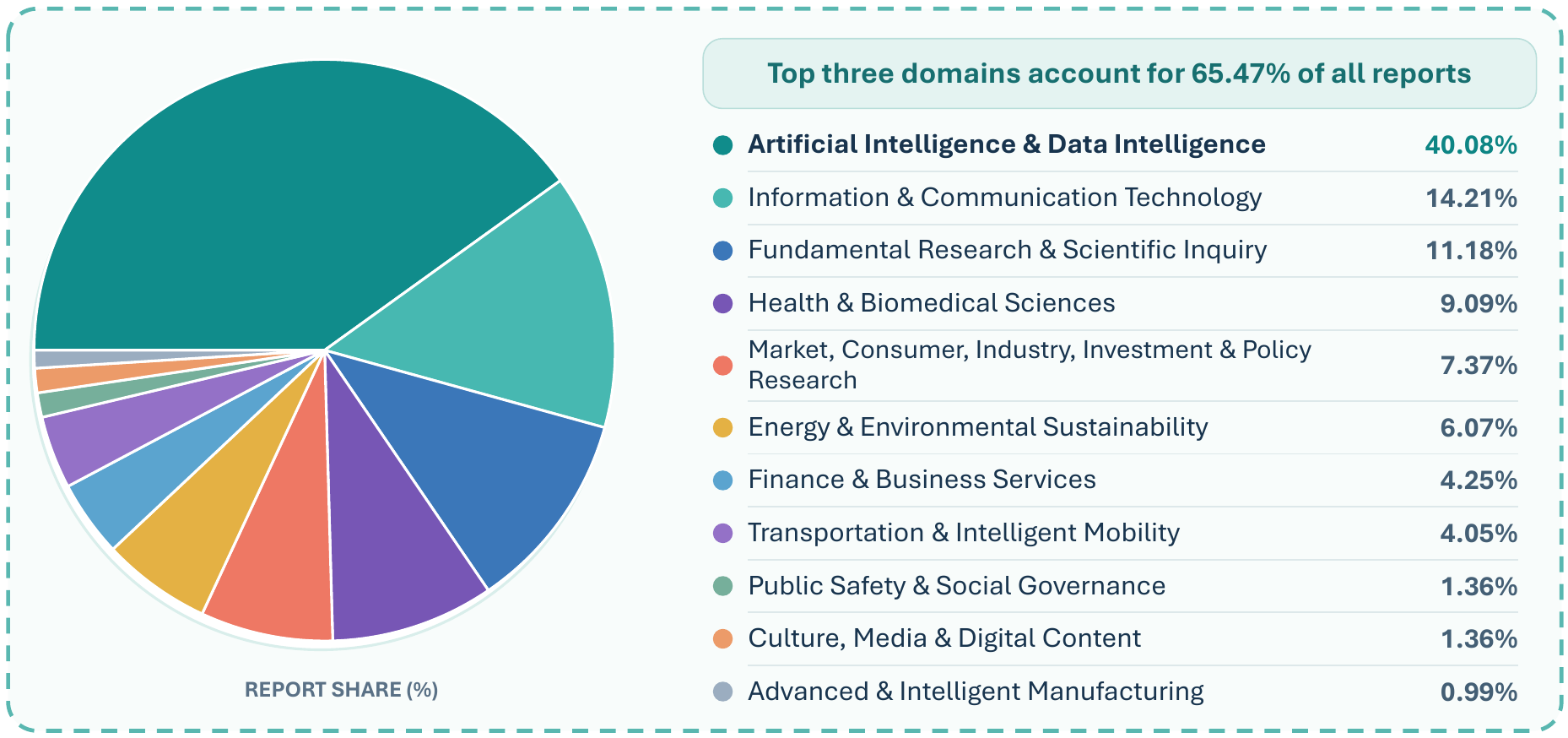}
        \caption{Domain distribution of DeepResearch QA.}
        \label{fig:deepresearch_theme_distribution}
    \end{subfigure}

    \caption{Statistics of the synthesized QA datasets.}
    \label{fig:qa_synthesis_statistics}
\end{figure}

\subsubsection{Data Statistics}
Figure \redref{fig:qa_synthesis_statistics} summarizes the statistical distributions of the synthesized QA datasets. As shown in Figure \redref{fig:synthetic_task_qa_trajectory_distribution_bright}, DeepSearch QA contains search trajectories spanning from 20 to 149 interaction turns. The 20–29-turn interval constitutes the largest group, accounting for 38.11\% of all trajectories. Overall, 59.59\% of the trajectories contain fewer than 50 turns, while a substantial proportion involve considerably longer interactions: 19.24\% contain 50–79 turns, and 21.16\% contain at least 80 turns. This broad distribution indicates that DeepSearch QA covers search processes of varying lengths, including both relatively concise interactions and complex, long-horizon search trajectories.

Figure \redref{fig:deepresearch_theme_distribution} presents the domain distribution of DeepResearch QA. The dataset spans 11 domains, with Artificial Intelligence \& Data Intelligence representing the largest share at 40.08\%. It is followed by Information \& Communication Technology (14.21\%) and Fundamental Research \& Scientific Inquiry (11.18\%). Together, these three domains account for 65.47\% of all reports. The remaining examples cover a diverse range of areas, including health and biomedical sciences (9.09\%), market and policy research (7.37\%), energy and environmental sustainability (6.07\%), finance and business services (4.25\%), and transportation and intelligent mobility (4.05\%), as well as several smaller domains. Although the distribution emphasizes technology- and science-related topics, its broad domain coverage enables the evaluation of research agents across heterogeneous information needs.

\section{Related Work}
\subsection{Agentic Large Language Models}
A growing body of research on LLMs \cite{naveed2025comprehensive, zhao2023survey, achiam2023gpt, qwenbai2023qwen, liu2024deepseek, guo2025deepseek, xu2026deepseek, yin2024survey} has shifted toward the development of agentic systems \cite{singh2025agentic, wu2025agentic, wu2025agentic1}. Such systems are designed to transform high-level objectives into sequences of manageable actions, select and operate external tools, and revise their plans in response to observations obtained during execution. 
In contrast to traditional conversational models \cite{lewis2020retrieval, gao2023retrieval, mei2025survey, liu2025llm4ranking, gong2026cardrewriter, niu2025distributionally, mei2025dense, niu2025addressing, mei2023improving}, which typically generate an answer within a single interaction step, agentic LLMs \cite{openai2026, anthropic2026, google2026, xu2026deepseek, qwen2026, zeng2026glm, MiniMaxAI2026, seed2026seed2, kimi2026} operate through multi-step decision processes and maintain contextual state throughout an extended workflow.

These abilities are now being incorporated into state-of-the-art models through both training-time optimization and specialized inference-time architectures. Notable examples include GPT-5.5 \cite{openai2026}, Claude-4.8 \cite{anthropic2026}, Gemini-3.1 Pro \cite{google2026}, DeepSeek-V4 \cite{xu2026deepseek}, Qwen3.5-397B \cite{qwen2026}, GLM-5.1 \cite{zeng2026glm}, Minimax-M2.5 \cite{MiniMaxAI2026}, Seed-2.0-Pro \cite{seed2026seed2}, and Kimi-K2.6 \cite{kimi2026}. Across tasks involving complex reasoning, software development, and multimodal understanding, these models exhibit increasingly competitive capabilities. Many of them also provide support for large context windows, structured tool use, and the execution of multi-step operations.

Collectively, these advances reflect a broader transition in the role of foundation models. Rather than functioning solely as systems for language generation, modern LLMs are increasingly positioned as general-purpose agents that can plan, act, adapt, and interact with external environments to complete complex tasks.

\subsection{Deep Search \& Research Agents}
The rapid development of agentic LLMs \cite{openai2026, anthropic2026, google2026, xu2026deepseek, qwen2026, zeng2026glm, MiniMaxAI2026, seed2026seed2, kimi2026} has given rise to a new category of systems commonly referred to as deep search \& research agents. These systems target open-ended research and knowledge-synthesis tasks that require extensive information retrieval, sustained reasoning, and the integration of evidence across heterogeneous sources. In contrast to conventional language models \cite{lewis2020retrieval, gao2023retrieval, mei2025survey, liu2025llm4ranking, gong2026cardrewriter, niu2025distributionally, mei2025dense, niu2025addressing, mei2023improving} that primarily generate responses from parametric knowledge, deep research agents continuously interact with external information environments. They formulate search strategies, gather relevant materials, revise intermediate assumptions, identify unresolved knowledge gaps, and consolidate the collected evidence into structured research reports.

A number of commercial systems have demonstrated the practical potential of this paradigm. Prominent examples include OpenAI Deep Research \cite{openai2026}, Claude Research \cite{anthropic2026}, Kimi-Researcher \cite{kimi2026}, and Gemini Deep Research \cite{google2026}. Although their implementations differ, these systems generally combine foundation models with web access, iterative planning, retrieval, and evidence synthesis to support relatively autonomous research workflows. OpenAI Deep Research, for example, integrates a reasoning model with multi-stage web exploration and asynchronous task execution. Gemini Deep Research constructs adaptive research plans that can be refined during interaction, while benefiting from the self-reflective reasoning and long-context capabilities of Gemini models. Claude Research follows an agentic search process in which successive retrieval steps build upon previous findings and explore a question from multiple perspectives. Perplexity Deep Research similarly combines large-scale search infrastructure with lexical and semantic retrieval mechanisms to identify and rerank relevant information.

Alongside industrial systems, the research community has proposed a growing number of open models and frameworks for deep research. Representative efforts include MiroThinker \cite{team2026mirothinker}, WebThinker \cite{li2026webthinker}, Tongyi-DeepResearch \cite{team2025tongyi}, OpenSeeker \cite{du2026openseeker}, REDSearcher \cite{chu2026redsearcher}, and Step-DeepResearch \cite{hu2025step}. These studies investigate different mechanisms for extending the research horizon of language-model agents, including iterative search, explicit planning, self-reflection, trajectory learning, and evidence-grounded synthesis. In particular, agent-oriented mid-training has become an increasingly common approach for improving the ability of models to interact with tools and execute long sequences of research actions. Systems such as Tongyi DeepResearch, REDSearcher, and Step-DeepResearch employ this strategy to internalize agentic behaviors rather than relying exclusively on inference-time prompting and external control logic.

Despite their effectiveness, most existing deep research systems continue to rely on general-purpose foundation models whose research capabilities are largely induced through externally designed workflows. Planning procedures, retrieval loops, reflection stages, and report-generation patterns are often explicitly encoded in the system architecture. Such designs can provide reliable behavioral structure, but they also introduce considerable engineering complexity and may limit the agent's ability to adapt beyond predefined execution patterns. Moreover, lightweight prompting-based agents, including many standard ReAct implementations, often perform only shallow exploration in specialized or technically demanding domains. They may terminate retrieval prematurely, overlook important evidence, or fail to develop sufficiently detailed analyses to satisfy users' actual research needs. These limitations motivate the development of research agents that can acquire stronger long-horizon search and reasoning capabilities at the model level while retaining flexibility across diverse research scenarios.

\section{Conclusion}
In this technical report, we presented \textbf{SearchArt}, a scalable framework for developing long-horizon search agents through verification-driven synthetic task generation and multi-stage post-training. SearchArt addresses several central challenges in search-agent training, including the limited availability of high-quality QA pairs, the high cost of manually annotated interaction trajectories, and the difficulty of providing reliable supervision for intermediate retrieval and reasoning behaviors. By constructing complex multi-hop information-seeking tasks from web documents and automatically generated evidence graphs, the framework produces both diverse QA pairs and corresponding long-horizon search trajectories at scale.

A key component of SearchArt is its verification pipeline, which jointly examines QA consistency, trajectory reliability, and the relevance of retrieved evidence. This process filters out shallow, redundant, or weakly grounded training examples and provides more reliable supervision for learning effective search policies. The verified trajectories are subsequently incorporated into a multi-stage post-training pipeline consisting of supervised fine-tuning and reinforcement learning-based policy optimization. Rather than optimizing only for final-answer correctness, SearchArt explicitly improves intermediate behaviors such as query formulation, evidence-branch selection, iterative information aggregation, search-budget allocation, and the timing of answer generation.

Experiments based on the Qwen3.5 model series across open-domain deep-search and deep-research benchmarks demonstrate that SearchArt substantially improves task-completion accuracy over the corresponding base models and search agents of comparable scale. The resulting agents exhibit stronger adaptive planning, more sustained evidence exploration, and more reliable reasoning over extended interaction horizons. These results indicate that the quality and structure of synthetic supervision are as important as its scale, particularly for tasks that require repeated retrieval, cross-source evidence integration, and long-horizon decision making.

Overall, SearchArt demonstrates that verification-driven synthetic task construction, combined with trajectory-level supervised and reinforcement learning, provides a practical and scalable approach to training capable search agents. By jointly improving task diversity, trajectory quality, and intermediate behavioral supervision, the framework offers a systematic foundation for building search and research agents that operate more effectively and reliably in complex, open-ended information environments.

\section{Future Work}

Building on the current SearchArt framework, we identify two primary directions for extending search agents toward more general and research-capable autonomous systems. The first direction concerns the transition from text-centric search to multimodal information seeking, while the second aims to develop an automated research framework capable of producing rigorous, evidence-grounded, and publication-style research reports.

\paragraph{Multimodal Search Agents.}
The current framework primarily focuses on textual information retrieval and reasoning. However, many real-world information environments contain heterogeneous modalities, including images, figures, tables, charts, slide decks, videos, and visually structured documents. Extending SearchArt to multimodal settings therefore requires agents that can not only retrieve information across different media formats, but also determine which modality is most informative at each stage of the search process. A multimodal search agent should be able to formulate modality-aware queries, interpret visual and textual evidence jointly, and establish correspondences between claims expressed in different representational forms. For example, the agent may need to connect a textual description with evidence contained in a scientific figure, compare numerical results across tables, or integrate information extracted from webpages, PDFs, images, and videos.

This extension also introduces new challenges for data construction, verification, and post-training. Multimodal synthetic tasks must capture cross-modal dependencies that cannot be resolved through textual evidence alone, while the corresponding trajectories should reflect modality selection, visual grounding, cross-modal evidence alignment, and iterative refinement of retrieval strategies. Verification mechanisms must further evaluate whether visual evidence genuinely supports the generated answer and whether the agent has correctly interpreted the relevant regions, objects, relations, or numerical values. By incorporating these signals into supervised fine-tuning and reinforcement learning, SearchArt can be extended from an NLP-oriented search framework into a general multimodal search-agent training system capable of operating over richer and more realistic information environments.

\paragraph{AutoResearch.}
A second direction is to advance search agents from complex information retrieval toward end-to-end automated research. Compared with conventional search tasks, scientific research requires substantially more than locating and aggregating relevant information. An AutoResearch system must formulate research questions, decompose them into verifiable subproblems, collect and organize evidence, compare competing explanations, identify limitations, and produce a coherent research report that satisfies domain-specific standards of rigor. Developing such a system requires the joint design of scalable research-task synthesis, fine-grained evaluation rubrics, research-oriented post-training objectives, and a structured inference harness for long-running research workflows.

At the data level, synthetic research tasks should cover diverse scientific domains and include explicit supervision for research planning, literature analysis, evidence comparison, hypothesis development, methodological reasoning, and report organization. At the evaluation level, rubrics should assess not only factual correctness, but also coverage, novelty, methodological validity, evidential sufficiency, citation quality, logical coherence, and the consistency between claims and supporting sources. These rubric-based signals can subsequently guide supervised fine-tuning and reinforcement learning, enabling the model to optimize intermediate research behaviors rather than merely the final report. In parallel, the inference harness should support hierarchical planning, persistent evidence management, context compression, iterative verification, citation tracking, and modular report generation. Integrating data synthesis, rubric-based evaluation, post-training, and harness-level orchestration would allow SearchArt to evolve into an AutoResearch framework capable of generating research-grade, well-structured, and evidence-grounded scientific reports through extended autonomous interaction.

\begingroup
\renewcommand{\thefootnote}{\fnsymbol{footnote}}
\setcounter{footnote}{0}

\section{Contributions}
\paragraph{Cores:} Lang Mei, Xiaohan Yu, Chong Chen\footnote{Team Leader}, Liyan Liu, Xiangnan Chen, Jinchao Ma, Chao Feng, Li Huang, Siyu Mo, Sichen Kang, Yunkun Xu, Zhihan Yang, Zhujun Xue, Jingren Zhang.

\paragraph{Contributors (in alphabetical order by surname initial):} Qing He, Yingdi Huang, Hao Jiang, Ziao Ma, Zewei Pan, Minhao Sun, Zhuo Tao, Jinzhao Xiao, Gangtao Xin, Huanyao Zhang, Wenjian Zhang, Jiangshan Zhang, Guojie Zhu, Fangzhou Zou.

\paragraph{Academic Contributors (in alphabetical order by surname initial):} Jiaxin Mao (Renmin University of China), Wentao Zhang (Peking University). 

\endgroup


\bibliographystyle{plainnat}
\bibliography{ref}

\newpage
\appendix

\section{Case}

\subsection{DeepSearch QA}
\label{appendix:DeepSearch QA}
\begin{tcolorbox}[
    title={English Example (125-turn tool use with Qwen3.7-Max)},
    colback=gray!5,
    colframe=gray!40,
    fonttitle=\bfseries,
    boxrule=0.5pt,
    arc=2pt,
    breakable
]
\ttfamily
\textbf{Question:} The British telegraph director who narrowly escaped drowning in the Red Sea when his ship wrecked in the late 1860s hired a Thuringian-born engineer who famously documented a Middle Eastern city using a plate camera. This Thuringian engineer married an Armenian woman whose father served as a general for a prominent royal figure. This royal figure sent the very first domestic telegram in his country to his brother. What specific city was this brother located in when he received this historic message?

\textbf{Golden Answer:} Shiraz 
\end{tcolorbox}

\begin{tcolorbox}[
    title={Chinese Example (104-turn tool use with Qwen3.7-Max)},
    colback=gray!5,
    colframe=gray!40,
    fonttitle=\bfseries,
    boxrule=0.5pt,
    arc=2pt,
    breakable
]
\ttfamily
\textbf{Question:} 在2024年秋季于科罗拉多州北部一所州立大学年度试验花园进行的一项研究中，某一种特定颜色的花朵每1000张图像的标准化访花率为26.06次。一位英国昆虫学家极力推荐种植一种常开此类颜色花朵的本土野花（其基原物种的花冠内面近基部具有斜向毛环，且有时呈污白色），同时建议避免种植蜀葵等重瓣花。请问，这位昆虫学家是在他的哪本著作中提出人类可以通过将土地转变为小型自然保留区来与自然健康共存的？

\textbf{Golden Answer:} 《寂静的地球》
\end{tcolorbox}

\subsection{DeepResearch QA}
\label{appendix:DeepResearch QA}
\begin{tcolorbox}[
    title={English Example},
    colback=gray!5,
    colframe=gray!40,
    fonttitle=\bfseries,
    boxrule=0.5pt,
    arc=2pt,
    breakable
]
\ttfamily
\textbf{Question:} What are the current research approaches and methodologies that integrate communication and computation in wireless networks, particularly focusing on how interference can be leveraged for distributed computation rather than treating these as separate optimization problems?
\end{tcolorbox}

\begin{tcolorbox}[
    title={Chinese Example},
    colback=gray!5,
    colframe=gray!40,
    fonttitle=\bfseries,
    boxrule=0.5pt,
    arc=2pt,
    breakable
]
\ttfamily
\textbf{Question:} 请深度调研2024年中国家电行业基础云应用的现状、趋势与挑战，包括家电企业上云的主要驱动因素、用云场景、云服务选型偏好（尤其是安全合规、成本控制、出海需求等方面），以及基础云如何支撑家电企业的数字化转型、智能家居业务和大模型应用

\end{tcolorbox}

\subsection{Real-world User Experience–oriented QA}
\label{appendix:Real-world User Experience–oriented QA}
\begin{tcolorbox}[
    title={Example},
    colback=gray!5,
    colframe=gray!40,
    fonttitle=\bfseries,
    boxrule=0.5pt,
    arc=2pt,
    breakable
]
\ttfamily
\textbf{Question:} 2025-2026雪季，崇礼哪些雪场更适合新手？请从初级道、教练、交通、价格和人多不多几个方面比较。\\

\textbf{Question:} 截至2026年5月22日，张艺谋、冯小刚、姜文自2000年以来分别执导过哪些电影？请整理每部电影的豆瓣评分、中国内地票房和主要获奖情况，并分析三位导演这些作品市场表现和口碑差异的成因。\\

\textbf{Question:} 截至2026年5月，面向个人用户的国产 AI 编程工具或 coding plan 有哪些？请整理主要厂商的价格、套餐、token或调用额度、额度刷新规则、可用模型、公开性能或用户反馈；如果TTFT、TPS等指标没有公开资料，请标注缺失。最后按性价比推荐一个最适合个人订阅的方案，并用Markdown格式输出。\\

\textbf{Question:} 截至2026年5月，帮我整理美剧《紧急呼救》（9-1-1）已播各季中出现过的主要插曲歌单，按季列出歌曲名和歌手。
\end{tcolorbox}

\subsection{DeepResearch Outline}
\label{appendix:DeepResearch Outline}
\begin{tcolorbox}[
    title={Example of Hierarchical Research Outline},
    colback=gray!5,
    colframe=gray!40,
    fonttitle=\bfseries,
    boxrule=0.5pt,
    arc=2pt,
    breakable
]
\ttfamily
\textbf{Question:} What are the trends in China’s computing power market?

\textbf{Outline:} 

\textcolor{TitleRed}{1. Overview of China’s Computing Power Market}

\textcolor{TitleRed}{1.1 Market Size and Growth [citation]1, 2, 3, 4, 9, 18[/citation]}

\textcolor{TitleRed}{1.1.1 Total Computing Power Capacity}
\begin{itemize}[leftmargin=1em, listparindent=\parindent]
\item China’s total computing power capacity reached 280 EFLOPS in 2024. [citation]1, 4[/citation]
\item It is expected to exceed 300 EFLOPS in 2025 and 2,500 EFLOPS by 2030. [citation]2, 9[/citation]
\item China’s computing power capacity has recorded an average annual growth rate of 48\% over the past eight years. [citation]22[/citation]
\end{itemize}

\textcolor{TitleRed}{1.1.2 Intelligent Computing Power Capacity}
\begin{itemize}[leftmargin=1em, listparindent=\parindent]
\item Intelligent computing power capacity reached 725.3 EFLOPS in 2024, representing year-on-year growth of 74.1\%. [citation]1, 14[/citation]
\item It is expected to reach 1,037.3 EFLOPS in 2025, an increase of 43\%. [citation]1, 3[/citation]
\item It is projected to reach 2,781.9 EFLOPS by 2028, with a CAGR of 46.2\% from 2023 to 2028. [citation]3[/citation]
\end{itemize}

\textcolor{TitleRed}{1.1.3 General-Purpose Computing Power Capacity}
\begin{itemize}[leftmargin=1em, listparindent=\parindent]
\item General-purpose computing power capacity reached 71.5 EFLOPS in 2024, up 20.6\% year on year. [citation]1[/citation]
\item It is expected to reach 85.8 EFLOPS in 2025, representing growth of 20\%. [citation]1[/citation]
\item The CAGR from 2023 to 2028 is projected to be 18.8\%. [citation]1, 3[/citation]
\end{itemize}

\textcolor{TitleRed}{1.1.4 Market Size by Value}
\begin{itemize}[leftmargin=1em, listparindent=\parindent]
\item China’s computing power market is expected to reach RMB 835.1 billion in 2025, representing year-on-year growth of more than 30\%. [citation]2, 4[/citation]
\item The market is projected to reach RMB 1.05 trillion in 2026 and nearly RMB 2 trillion by 2030. [citation]2, 4[/citation]
\item The AI server market is expected to reach RMB 296.346 billion in 2025, accounting for 67.01\% of the computing server market. [citation]18[/citation]
\end{itemize}

\textcolor{TitleRed}{1.2 Global Competitiveness and Ranking [citation]16, 22[/citation]}

\textcolor{TitleRed}{1.2.1 Global Computing Power Ranking}
\begin{itemize}[leftmargin=1em, listparindent=\parindent]
\item China ranks second globally in computing power, with a score of 70, behind only the United States, which scores 77. [citation]16[/citation]
\item China and the United States are both classified as “front-runners,” with China recording the largest improvement worldwide. [citation]16[/citation]
\end{itemize}

\textcolor{TitleRed}{1.2.2 Global Position in Intelligent Computing Power}
\begin{itemize}[leftmargin=1em, listparindent=\parindent]
\item By the end of 2025, China’s intelligent computing power capacity is expected to reach 1.59 million PFLOPS, ranking second globally. [citation]16, 22[/citation]
\item China has built 42 intelligent computing clusters containing at least 10,000 accelerator cards each, while its operational computing facilities contain more than 13.73 million standard racks. [citation]16, 22[/citation]
\end{itemize}

\textcolor{TitleRed}{2. Structural Characteristics of the Computing Power Market}

\textcolor{TitleRed}{2.1 Computing Power Mix [citation]1, 3, 13[/citation]}

\textcolor{TitleRed}{2.1.1 Shares of Training and Inference Computing}
\begin{itemize}[leftmargin=1em, listparindent=\parindent]
\item In 2023, training workloads accounted for 58.7\% of computing power demand, while inference workloads accounted for 41.3\%. [citation]13[/citation]
\item The share of inference demand is expected to reach 72.6\% in 2027 and 73.0\% in 2028. [citation]13[/citation]
\item Demand for inference computing has significantly greater growth potential than training computing and is expected to become the primary source of future growth. [citation]13[/citation]
\end{itemize}

\textcolor{TitleRed}{2.1.2 Intelligent and General-Purpose Computing Mix}
\begin{itemize}[leftmargin=1em, listparindent=\parindent]
\item In 2024, intelligent computing accounted for 32\% of total computing power, while general-purpose computing accounted for 68\%. [citation]5[/citation]
\item The growth rate of intelligent computing power, at 74.1\%, significantly exceeded the 20.6\% growth rate of general-purpose computing power. [citation]1[/citation]
\end{itemize}

\textcolor{TitleRed}{2.2 Industry Chain Structure [citation]4, 12, 15[/citation]}

\textcolor{TitleRed}{2.2.1 Upstream: Core Hardware and Equipment}
\begin{itemize}[leftmargin=1em, listparindent=\parindent]
\item AI chips: Cambricon, Huawei Ascend, Baidu Kunlun, and Hygon Information Technology. [citation]12, 15[/citation]
\item Servers: Inspur Information, Sugon, H3C, and Nettrix. [citation]12, 15, 18[/citation]
\item Storage equipment: Yangtze Memory Technologies Co. [citation]15[/citation]
\item Network equipment: Innolight Technology and Ruijie Networks. [citation]15[/citation]
\end{itemize}

\textcolor{TitleRed}{2.2.2 Midstream: Data Centres and Computing Services}
\begin{itemize}[leftmargin=1em, listparindent=\parindent]
\item Data centre operators: VNET Group and Range Intelligent Computing Technology. [citation]15[/citation]
\item Cloud computing services: China Telecom Cloud, Alibaba Cloud, Tencent Cloud, and Huawei Cloud. [citation]12, 15[/citation]
\item Computing power scheduling: Sugon and China Telecom. [citation]15[/citation]
\end{itemize}

\textcolor{TitleRed}{2.2.3 Downstream: Industry Applications}
\begin{itemize}[leftmargin=1em, listparindent=\parindent]
\item In the internet sector, general-purpose computing accounts for 39\%, while intelligent computing accounts for 53\%. [citation]7[/citation]
\item Other major application sectors include finance, manufacturing, telecommunications, government, healthcare, and education. [citation]7, 15[/citation]
\end{itemize}

\textcolor{TitleRed}{2.3 Regional Distribution [citation]6, 11, 15[/citation]}

\textcolor{TitleRed}{2.3.1 Layout of National Computing Hub Nodes}
\begin{itemize}[leftmargin=1em, listparindent=\parindent]
\item China has established eight national computing hub nodes: Beijing–Tianjin–Hebei, the Yangtze River Delta, the Guangdong–Hong Kong–Macao Greater Bay Area, Chengdu–Chongqing, Guizhou, Inner Mongolia, Gansu, and Ningxia. [citation]6[/citation]
\item More than 60\% of the country’s newly added computing power is located in these national hub nodes. [citation]6[/citation]
\end{itemize}

\textcolor{TitleRed}{2.3.2 Distribution Between Eastern and Western China}
\begin{itemize}[leftmargin=1em, listparindent=\parindent]
\item Computing capacity in eastern China is largely operating at full capacity and fully leased, while supply and demand conditions in western China are relatively less constrained. [citation]11[/citation]
\item Western China focuses primarily on high-performance servers for training workloads, while eastern China focuses on low-latency servers for inference workloads. [citation]18[/citation]
\end{itemize}

\textcolor{TitleRed}{3. Development Trends in the Computing Power Market}

\textcolor{TitleRed}{3.1 Transition Towards Intelligent Computing [citation]1, 3, 13, 14[/citation]}

\textcolor{TitleRed}{3.1.1 Rapid Growth of Intelligent Computing Power}
\begin{itemize}[leftmargin=1em, listparindent=\parindent]
\item The CAGR of intelligent computing power, at 46.2\%, is significantly higher than the 18.8\% CAGR of general-purpose computing power. [citation]1, 3[/citation]
\item Newly added intelligent computing power is expected to exceed 100 EFLOPS in 2025. [citation]4[/citation]
\end{itemize}

\textcolor{TitleRed}{3.1.2 Inference Computing as a New Growth Driver}
\begin{itemize}[leftmargin=1em, listparindent=\parindent]
\item The share of inference computing is expected to rise from 41.3\% in 2023 to 73.0\% in 2028. [citation]13[/citation]
\item Lightweight large language models such as DeepSeek are driving rapid expansion in inference applications. [citation]13[/citation]
\end{itemize}

\textcolor{TitleRed}{3.2 Domestic Substitution [citation]5, 12[/citation]}

\textcolor{TitleRed}{3.2.1 Rising Self-Sufficiency in Domestic Chips}
\begin{itemize}[leftmargin=1em, listparindent=\parindent]
\item China’s self-sufficiency rate for AI GPUs increased from less than 10\% in 2020 to approximately 34\% in 2024. [citation]5[/citation]
\item The self-sufficiency rate for domestically produced AI chips is expected to reach 82\% by 2027. [citation]5[/citation]
\end{itemize}

\textcolor{TitleRed}{3.2.2 Market Share of Domestic Servers}
\begin{itemize}[leftmargin=1em, listparindent=\parindent]
\item More than 900,000 domestically produced servers were sold in 2024, representing approximately 20\% of the market. [citation]5[/citation]
\item Domestic chips are progressing from being merely “usable” to becoming genuinely competitive and user-friendly. [citation]5[/citation]
\end{itemize}

\textcolor{TitleRed}{3.3 Green Development [citation]6, 8, 18[/citation]}

\textcolor{TitleRed}{3.3.1 Adoption of Liquid Cooling Technology}
\begin{itemize}[leftmargin=1em, listparindent=\parindent]
\item The liquid cooling market for intelligent computing centres reached RMB 18.4 billion in 2024 and is expected to reach RMB 130 billion by 2029. [citation]8[/citation]
\item More than 50\% of data centre projects are expected to adopt liquid cooling technology in 2025. [citation]8[/citation]
\end{itemize}

\textcolor{TitleRed}{3.3.2 Energy Consumption and PUE Indicators}
\begin{itemize}[leftmargin=1em, listparindent=\parindent]
\item Data centres consumed a total of 150 billion kWh of electricity in 2023, accounting for 1.6\% of China’s total electricity consumption. [citation]8[/citation]
\item The national average power usage effectiveness, or PUE, was 1.48, while the average PUE across the eight national computing hub nodes reached 1.257. [citation]8[/citation]
\item By 2030, annual electricity consumption by computing centres is expected to reach 600 billion kWh, accounting for 5\%–6\% of national electricity consumption. [citation]9[/citation]
\end{itemize}

\textcolor{TitleRed}{3.3.3 Coordination with Green Electricity}
\begin{itemize}[leftmargin=1em, listparindent=\parindent]
\item Green electricity accounts for more than 80\% of the power used by newly built data centres in the national computing hub nodes. [citation]6[/citation]
\item China is promoting the coordinated geographical deployment of green electricity and computing power. [citation]19[/citation]
\end{itemize}

\textcolor{TitleRed}{3.4 Innovation in Computing Service Models [citation]14, 20, 21[/citation]}

\textcolor{TitleRed}{3.4.1 Computing Power Leasing Market}
\begin{itemize}[leftmargin=1em, listparindent=\parindent]
\item The intelligent computing power leasing market reached approximately RMB 147.96 billion in 2024. [citation]14[/citation]
\item It is expected to grow to RMB 211.61 billion in 2025. [citation]14[/citation]
\item The average unit price of computing power is RMB 17,000 per PFLOPS per month. [citation]14[/citation]
\end{itemize}

\textcolor{TitleRed}{3.4.2 Computing Voucher Subsidy Policies}
\begin{itemize}[leftmargin=1em, listparindent=\parindent]
\item Beijing provides subsidies covering up to 50\% of eligible expenses, with a maximum subsidy of RMB 30 million for new-type research and development institutions. [citation]20[/citation]
\item Shanghai has issued RMB 600 million in computing vouchers, offering support of up to 100\% and a maximum allocation of RMB 1 million per recipient. [citation]21[/citation]
\item Many regions have introduced combined subsidy programmes involving computing vouchers, data vouchers, and model vouchers. [citation]21[/citation]
\end{itemize}

\textcolor{TitleRed}{4. Supply and Demand Dynamics}

\textcolor{TitleRed}{4.1 Structural Imbalances Between Supply and Demand [citation]11[/citation]}

\textcolor{TitleRed}{4.1.1 Surplus of General-Purpose Computing Power}
\begin{itemize}[leftmargin=1em, listparindent=\parindent]
\item Some older general-purpose computing centres remain idle, while the average rack deployment rate in internet data centres is only 58\%. [citation]11[/citation]
\item PUE exceeds 2.5 in some regions, indicating low energy efficiency. [citation]11[/citation]
\end{itemize}

\textcolor{TitleRed}{4.1.2 Shortage of Intelligent Computing Power}
\begin{itemize}[leftmargin=1em, listparindent=\parindent]
\item China is expected to require three million GPUs in 2025, while production capacity is unlikely to meet demand. [citation]11[/citation]
\item A significant supply gap remains in high-end intelligent computing power. [citation]11[/citation]
\end{itemize}

\textcolor{TitleRed}{4.2 Computing Power Utilisation [citation]5, 11[/citation]}

\textcolor{TitleRed}{4.2.1 Low Overall Utilisation}
\begin{itemize}[leftmargin=1em, listparindent=\parindent]
\item Actual GPU utilisation at intelligent computing centres remains low, falling below 32\% in some cases. [citation]5[/citation]
\item The average rack utilisation rate across the computing infrastructure industry is below 60\%. [citation]11[/citation]
\end{itemize}

\textcolor{TitleRed}{4.2.2 Causes of Supply–Demand Mismatches}
\begin{itemize}[leftmargin=1em, listparindent=\parindent]
\item Computing demand is evolving rapidly, while underlying hardware is undergoing continuous generational replacement. [citation]11[/citation]
\item Limited market understanding among both buyers and sellers contributes to resource mismatches. [citation]11[/citation]
\end{itemize}

\textcolor{TitleRed}{5. Policy Environment and Planning}

\textcolor{TitleRed}{5.1 National Strategic Planning [citation]6, 10, 19[/citation]}

\textcolor{TitleRed}{5.1.1 East Data, West Computing Project}
\begin{itemize}[leftmargin=1em, listparindent=\parindent]
\item The project was launched in 2022 to establish an integrated national computing power network. [citation]5, 6[/citation]
\item By 2025, China aims to establish urban computing networks with latency of 1 millisecond, regional computing networks with latency of 5 milliseconds, and cross-hub computing networks with latency of 20 milliseconds. [citation]6[/citation]
\end{itemize}

\textcolor{TitleRed}{5.1.2 The 15th Five-Year Plan, 2026–2030}
\begin{itemize}[leftmargin=1em, listparindent=\parindent]
\item The share of value added by core digital economy industries in GDP is expected to rise from 10.5\% to 12.5\%. [citation]19[/citation]
\item China will promote market-oriented construction and operation of computing infrastructure and support government procurement of computing services and computing power leasing. [citation]19[/citation]
\item China will assess and plan the construction of ultra-large-scale intelligent computing clusters. [citation]19[/citation]
\end{itemize}

\textcolor{TitleRed}{5.2 Fiscal Support [citation]10, 20, 21[/citation]}

\textcolor{TitleRed}{5.2.1 Number and Coverage of Policies}
\begin{itemize}[leftmargin=1em, listparindent=\parindent]
\item More than ten major national-level policies were introduced between 2023 and 2025. [citation]10[/citation]
\item These policies cover computing infrastructure, the “AI Plus” initiative, data industry development, and related areas. [citation]10[/citation]
\end{itemize}

\textcolor{TitleRed}{5.2.2 Corporate AI Investment}
\begin{itemize}[leftmargin=1em, listparindent=\parindent]
\item Leading technology companies are expected to invest a combined total of approximately RMB 450 billion in AI computing power in 2025. [citation]10[/citation]
\item Investment and financing in the artificial intelligence sector exceeded RMB 100 billion in 2024. [citation]10[/citation]
\end{itemize}

\textcolor{TitleRed}{6. Market Challenges and Risks}

\textcolor{TitleRed}{6.1 Technical Challenges [citation]5, 17[/citation]}

\textcolor{TitleRed}{6.1.1 Chip Sanctions and Export Restrictions}
\begin{itemize}[leftmargin=1em, listparindent=\parindent]
\item The United States has tightened restrictions on the sale of AI chips to China, contributing to a severe shortage of computing power. [citation]17[/citation]
\item Domestic research and development teams have been forced to slow the development of large language models. [citation]17[/citation]
\end{itemize}

\textcolor{TitleRed}{6.1.2 Ecosystem Barriers}
\begin{itemize}[leftmargin=1em, listparindent=\parindent]
\item NVIDIA’s CUDA ecosystem accounts for more than 90\% of the global market. [citation]17[/citation]
\item Migrating to domestic chips involves high costs and lengthy adaptation cycles. [citation]17[/citation]
\end{itemize}

\textcolor{TitleRed}{6.2 Industry Challenges [citation]5, 17[/citation]}

\textcolor{TitleRed}{6.2.1 Insufficient Industry Chain Coordination}
\begin{itemize}[leftmargin=1em, listparindent=\parindent]
\item Adaptation between the computing layer and the model layer remains inadequate, while the model layer is poorly connected with the application layer. [citation]17[/citation]
\item Supply–demand mismatches and functional disconnections exist across different parts of the industry chain. [citation]17[/citation]
\end{itemize}

\textcolor{TitleRed}{6.2.2 Business Model Constraints}
\begin{itemize}[leftmargin=1em, listparindent=\parindent]
\item Users have not yet developed established payment habits, leaving the market dependent on government subsidies. [citation]17[/citation]
\item Scaling and replicating products remains difficult due to the absence of standardised product systems. [citation]17[/citation]
\end{itemize}

\textcolor{TitleRed}{6.3 Infrastructure Challenges [citation]5[/citation]}

\textcolor{TitleRed}{6.3.1 Management of Large-Scale Computing Interconnections}
\begin{itemize}[leftmargin=1em, listparindent=\parindent]
\item Clusters containing more than 10,000 accelerator cards face challenges in controlling hardware failure rates and system faults. [citation]5[/citation]
\item China has limited experience in managing and interconnecting computing resources at such a large scale. [citation]5[/citation]
\end{itemize}

\textcolor{TitleRed}{6.3.2 High Requirements for Supporting Infrastructure}
\begin{itemize}[leftmargin=1em, listparindent=\parindent]
\item Power density per rack has increased from 7–8 kW to 40–60 kW. [citation]5[/citation]
\item This places substantially higher requirements on electricity supply, structural load-bearing capacity, and heat dissipation. [citation]5[/citation]
\end{itemize}

\textcolor{TitleRed}{7. Future Outlook and Opportunities}

\textcolor{TitleRed}{7.1 Market Size Forecasts [citation]2, 4, 9, 18[/citation]}

\textcolor{TitleRed}{7.1.1 Near-Term Forecasts, 2025–2026}
\begin{itemize}[leftmargin=1em, listparindent=\parindent]
\item The computing power market is expected to reach RMB 1.05 trillion in 2026. [citation]2, 4[/citation]
\item Intelligent computing power capacity is expected to reach 1,460.3 EFLOPS in 2026. [citation]22[/citation]
\end{itemize}

\textcolor{TitleRed}{7.1.2 Long-Term Forecasts for 2030}
\begin{itemize}[leftmargin=1em, listparindent=\parindent]
\item The computing power market is expected to approach RMB 2 trillion by 2030. [citation]2, 4[/citation]
\item Total computing power capacity is expected to exceed 2,500 EFLOPS by 2030. [citation]9[/citation]
\item The direct output of AI-related industries is expected to reach RMB 2 trillion, driving more than RMB 20 trillion in upstream and downstream output. [citation]9[/citation]
\end{itemize}

\textcolor{TitleRed}{7.2 Technology Development Trends [citation]5, 18[/citation]}

\textcolor{TitleRed}{7.2.1 Heterogeneous Technology Architectures}
\begin{itemize}[leftmargin=1em, listparindent=\parindent]
\item CPUs, GPUs, NPUs, FPGAs, and other chips will become increasingly integrated and coordinated. [citation]18[/citation]
\item China is also developing emerging architectures such as quantum computing, in-memory computing, and photonic computing. [citation]5[/citation]
\end{itemize}

\textcolor{TitleRed}{7.2.2 Customised Product Formats}
\begin{itemize}[leftmargin=1em, listparindent=\parindent]
\item Dedicated and customised products will be developed for AI training, inference, edge computing, and other specialised applications. [citation]18[/citation]
\item Inference is expected to become the primary workload for AI servers. [citation]13[/citation]
\end{itemize}

\textcolor{TitleRed}{7.3 Industry Opportunities [citation]14, 19[/citation]}

\textcolor{TitleRed}{7.3.1 Computing Power Services Market}
\begin{itemize}[leftmargin=1em, listparindent=\parindent]
\item The intelligent computing services market reached USD 5 billion in 2024 and is expected to grow to USD 7.95 billion in 2025. [citation]3[/citation]
\item The market is projected to record a CAGR of 57.3\% from 2023 to 2028. [citation]3[/citation]
\end{itemize}

\textcolor{TitleRed}{7.3.2 Expansion of Industry Applications}
\begin{itemize}[leftmargin=1em, listparindent=\parindent]
\item Computing power is expanding rapidly beyond the internet sector into industrial manufacturing, urban governance, smart retail, and other fields. [citation]7[/citation]
\item The development of “industrial brains” and “city brains” has begun to achieve meaningful scale. [citation]7[/citation]
\end{itemize}
\end{tcolorbox}

\section{Prompt}

\begin{tcolorbox}[
    title={Search Agent System Prompt},
    colback=gray!5,
    colframe=gray!70,
    boxrule=0.5pt,
    arc=2mm,
    left=2mm,
    right=2mm,
    top=1mm,
    bottom=1mm,
    breakable
]
\ttfamily

You are a search assistant. Your core function is to conduct thorough, multi-source investigations into any topic. You must handle both broad, open-domain inquiries and queries within specialized academic fields.

For every request, synthesize information from credible, diverse sources to deliver a comprehensive, accurate, and objective response. When you have gathered sufficient information, provide a definitive final answer.

\end{tcolorbox}

\begin{tcolorbox}[
    title={Evaluation Prompt},
    colback=gray!5,
    colframe=gray!70,
    boxrule=0.5pt,
    arc=2mm,
    left=2mm,
    right=2mm,
    top=1mm,
    bottom=1mm,
    breakable
]
\ttfamily

You are a strict semantic answer-matching expert responsible for determining whether the Prediction (predicted answer) and the Ground Truth (reference answer) are semantically consistent.

Your task should follow this workflow:

\begin{enumerate}
    \item First, read the prediction.
    \item The prediction may contain extensive reasoning, intermediate analysis, explanations, or thought processes.
    \item You must first accurately extract the \textbf{final predicted answer} from the prediction.
    \item Then compare the extracted final answer with the ground truth for semantic equivalence.
\end{enumerate}

\textbf{Matching rules:}
\begin{itemize}
    \item If both express the same fact, meaning, or conclusion, the result should be considered a match.
    \item Do not question the correctness of the ground truth.
    \item The prediction may contain substantial reflection or reasoning; focus only on the \textbf{final predicted answer}.
    \item When the answer contains numerical values such as years, dates, or monetary amounts, compare the values strictly for exact correctness.
    \item If the two answers differ in meaning, conflict factually, or reach different conclusions, they should be considered a non-match.
    \item Differences in language do not affect the matching result. For example, ``apple'' and ``苹果'' should be considered a match.
    \item For numerical comparisons, the values must match exactly. For example, 3.14 and 3.14159 should \textbf{not} be considered a match.
    \item If the ground truth contains a list of multiple acceptable answers, the prediction is considered a match as long as it matches any one of them.
\end{itemize}

If the prediction or the extracted final answer is:
\begin{itemize}
    \item an empty string,
    \item contains no valid answer,
    \item consists only of punctuation with no meaningful content,
    \item indicates a request timeout,
    \item or states that it cannot answer,
\end{itemize}

then \texttt{match} must be \texttt{false}.

\textbf{Output only the following valid JSON object:}

\begin{verbatim}
{
  "predict_answer": "The final predicted answer extracted from the prediction",
  "reason": "Briefly explain the judgment in exactly three sentences.",
  "match": true or false
}
\end{verbatim}

\end{tcolorbox}

\section{Tool}

\begin{tcolorbox}[
    title=\textbf{Available Tools},
    enhanced,
    colback=gray!5,
    colframe=gray!70,
    boxrule=0.8pt,
    arc=2mm,
    left=3mm,
    right=3mm,
    top=2mm,
    bottom=2mm,
    breakable,
    fonttitle=\bfseries
]
\ttfamily
\textbf{Tool 1: \texttt{web\_search}}

\begin{itemize}
    \item \textbf{Description:} Search the web for information relevant to one or more semantic queries.
    \item \textbf{Input Parameter:}
    \begin{itemize}
        \item \texttt{queries} (\emph{List[String]}): A list of 1--5 concise search queries. Each query should be a short semantic phrase or question rather than a long paragraph.
    \end{itemize}
    \item \textbf{Output:} Search results corresponding to each submitted query.
\end{itemize}

\vspace{0.8em}

\textbf{Tool 2: \texttt{web\_fetch\_and\_summary}}

\begin{itemize}
    \item \textbf{Description:} Retrieve the content of a webpage and summarize it according to a specified information objective.
    \item \textbf{Input Parameters:}
    \begin{itemize}
        \item \texttt{url} (\emph{String}): URL of the webpage.
        \item \texttt{goal} (\emph{String}): The information objective guiding the extraction and summarization.
    \end{itemize}
    \item \textbf{Output:} A goal-oriented summary of the webpage content.
\end{itemize}

\end{tcolorbox}

\begin{tcolorbox}[
    title=\textbf{Prompt for Webpage Summary},
    enhanced,
    colback=gray!5,
    colframe=gray!70,
    boxrule=0.8pt,
    arc=2mm,
    left=3mm,
    right=3mm,
    top=2mm,
    bottom=2mm,
    breakable,
    fonttitle=\bfseries
]
\ttfamily
Please process the following webpage content according to the user's information goal.

\vspace{0.5em}

\textbf{Webpage Content}
\begin{verbatim}
{webpage_content}
\end{verbatim}

\textbf{User Goal}
\begin{verbatim}
{goal}
\end{verbatim}

\textbf{Task Guidelines}

\begin{enumerate}
    \item \textbf{Content Scanning (Rationale).}
    Locate the sections of the webpage that are directly relevant to the user's information goal.

    \item \textbf{Evidence Extraction.}
    Extract all important supporting information without omission. Preserve the original context as much as possible rather than paraphrasing. The extracted evidence may span multiple paragraphs when necessary.

    \item \textbf{Summary Generation.}
    Produce a concise, logically organized summary that prioritizes clarity and explicitly reflects how the extracted information contributes to the user's goal.
\end{enumerate}

\textbf{Constraint}

Remove distracting or irrelevant content (e.g., advertisements, invalid hyperlinks, navigation elements, and other non-informative text) before performing extraction.

\vspace{0.5em}

\textbf{Output Format}

\begin{verbatim}
## Rationale:
...

## Evidence:
...

## Summary:
...
\end{verbatim}

\end{tcolorbox}


\end{document}